\documentclass[12pt,reqno]{amsart}
\usepackage{xcolor}
\usepackage{fullpage}
\usepackage{amsfonts}
\usepackage{amsmath}
\usepackage{mathbbol}
\usepackage{amssymb}
\usepackage{times}
\usepackage{graphicx}
\usepackage{float} 

\usepackage{mathtools}
\usepackage{multirow}
\usepackage{enumerate}

\usepackage{epsfig}

\newtheorem{theorem}{Theorem}[section]

\newtheorem{proposition}[theorem]{Proposition}
\newtheorem{lemma}[theorem]{Lemma}

\newcommand{\R}{\mathbb R}

\newcommand{\myC}{\mathbb C}

\renewcommand{\det}{{\,\rm det}\:}

\newcommand{\Tr}{{\,\rm Tr}\:}

\renewcommand{\exp}[1]{\mathrm{exp} \left\{#1\right\}}

\newcommand{\de}[1]{\mathrm{d}\, #1}

\def\sumd{\displaystyle\sum}

\def\intd{\displaystyle\int}

\def\prodd{\displaystyle\prod}

\DeclareMathOperator*\diag{diag}

\DeclareMathOperator*\im{Im}
\DeclareMathOperator*\re{Re}

\begin{document}
	
\sloppy

%


\newtheorem*{theorem*}{Theorem}
\newtheorem*{proposition*}{Theorem}
\newtheorem*{lemma*}{Theorem}

\newcommand{\tex}{{\tt .tex}}

\renewcommand*{\vec}[1]{\boldsymbol{#1}}

\title[Article]
{ 
{ Extreme eigenvalues and the emerging outlier in rank-one non-Hermitian deformations of the Gaussian Unitary Ensemble}}

\author{Yan V. Fyodorov$^{1}$, Boris A. Khoruzhenko$^{2}$ and Mihail Poplavskyi $^{2}$}
\address{$^1$ Department of Mathematics, King's College London, London  WC2R 2LS, United Kingdom}
\address{$^2$ School of Mathematical Sciences, Queen Mary University of London, London E1 4NS, United Kingdom}


\begin{abstract}
Complex eigenvalues of random matrices  $J=\text{GUE }+ i\gamma \diag (1, 0, \ldots, 0)$ provide the simplest model for studying resonances in wave scattering from a quantum chaotic system via a single open channel. It is known that in the limit of large matrix dimensions $N\gg 1$ the eigenvalue density of $J$  undergoes an abrupt restructuring at $\gamma = 1$, the critical threshold beyond which a single eigenvalue outlier (``broad resonance'') appears.  We provide a detailed description of this restructuring transition, including the scaling with $N$ of the width of the critical region about the outlier threshold  $\gamma=1$ and the associated scaling for the real parts (``resonance positions'') and imaginary parts (``resonance widths'') of
the eigenvalues which are farthest away from the real axis.
In the critical regime we determine the density of such extreme eigenvalues, and show how the outlier gradually separates itself from the rest of the extreme eigenvalues. Finally, we describe the fluctuations in the height
of the eigenvalue outlier for large but finite $N$ in terms of the associated large deviation function.
\end{abstract}

\maketitle

\section{Introduction}

Rank-one non-normal deformations of the Gaussian and Circular Unitary Ensembles are a useful analytic tool for studying statistics of resonances in quantum scattering from a chaotic domain via a single channel \cite{FyodSomm97,FyodSommRev03}.  As surveyed in \cite{FyodSommRev03,F2022}, these random matrix ensembles are integrable in the sense that the joint probability density of their complex eigenvalues and, in some spectral scaling limits of interest, the eigenvalue correlation functions can be determined in a closed form.  
Such integrability, which also proves to be useful in other physics contexts, see e.g. \cite{PS2018}, extends  to a certain degree to the deformed $\beta-$Gaussian and $\beta-$circular ensembles \cite{Kozhan17,KK2017}, especially to the classical values $\beta=1,4$ \cite{FO2022,FOTunpub}, but is lost if the underlying normal random matrix ensemble (Hermitian or unitary) is not integrable, as is the case with, e.g., finite rank non-Hermitian deformations of Wigner matrices \cite{RW2017,outlier,DubErd2021} or band matrices \cite{MSTS2021}. Still, the latter matrices are found to share, in appropriate parameter ranges, some statistical characteristics of their complex eigenvalues and eigenvectors with their integrable counterparts.

In this paper we aim to investigate complex eigenvalues with extreme imaginary parts for the rank-one non-Hermitian deformations of the Gaussian Unitary Ensemble (GUE) by exploiting the above-mentioned integrability. The latter feature gives access to the asymptotics of the eigenvalue density in the complex plane on mesoscopic scales and allows us to carry out a quantitative analysis of the separation of the eigenvalue outlier (which is known to exist in this model \cite{RW2017,outlier}) from the rest of the eigenvalues. Eigenvalue outliers in the complex plane have recently attracted renewed interest \cite{T2013,RR2014,FI2019,DubErd2021}. Our analysis refines and complements the existing knowledge about the outliers of nearly Hermitian matrices \cite{RW2017,outlier,DubErd2021}  albeit for arguably the simplest model of its type. As we will demonstrate, despite the simplicity of the model, its extreme eigenvalues exhibit an interesting transition at a certain value of the deformation parameter, with rich critical behaviour which deserves to be studied in more detail.

 The non-Hermitian matrices that we consider are of the form
\begin{align}\label{defens}
	J = H+i \Gamma, 
\end{align}
where $H$ 
is a GUE matrix and $\Gamma$ is a diagonal matrix with all diagonal entries being zero except the first one,
\begin{align}
\label{defens_Gamma}
\Gamma= \gamma \diag (1, 0, \ldots, 0). 
\end{align}
Denoting the matrix dimension by $N$, we fix the global spectral scale by the condition that  the expected value of  $\Tr H^2$ is $N$.
Then the joint probability density function (JDPF) of matrix elements of the GUE matrix $H$ is
\begin{align}\label{GUE}
f_N(H) = const \times \exp{-\frac{N}{2}\Tr H^2}.
\end{align}
With this normalisation, the limiting eigenvalue distribution of $H$, as the matrix dimension is approaching infinity, is supported on the interval $[-2,2]$, and, inside this interval,
the eigenvalue density  is $\nu(X) = \frac{1}{2\pi}\sqrt{4-X^2}$.

Note that due to the invariance of the JPDF (\ref{GUE}) with respect to unitary rotations $H\to UHU^{-1}$ one may equivalently replace $\Gamma$ in (\ref{defens_Gamma}) with any other rank-one Hermitian matrix.
Without loss of generality we may also assume  $\gamma $ to be positive.
Then
the eigenvalues $X_j+iY_j$ of
matrices $J$ \eqref{defens} -- \eqref{GUE}
are all in the upper half of the complex plane
and for $N$ large they all, except possibly one outlier, lie just above the interval $[-2,2]$ of the real line.  Whether
such an outlier is present
or not is determined by the value of $\gamma$.
For fixed values of
$\gamma <1$,
almost surely,
for $N$ sufficiently large,
 \emph{all} $N$ eigenvalues 
lie within distance $c_N N^{-1}$ from the real line, with $c_N =o(N^{\epsilon})$ for every $\epsilon>0$ \cite{RW2017}.
And if $\gamma>1$ then the same is true of all but one eigenvalue. 
This outlier lies much higher in the complex plane: to leading order in $N$, its imaginary part (the  ``height'')  is $\gamma - \gamma^{-1}$ \cite{RR2014, RW2017, outlier}.
For precise statements and proofs we refer the reader to \cite{RW2017, outlier} where these and similar facts were established
for finite rank non-Hermitian deformations of real symmetric matrices with independent matrix entries.

For finite but large matrix dimensions, one would expect to find a transition region of  infinitesimal
width
$\Omega$
about the outlier threshold value  $\gamma=1$ which captures the emergence of the outlier from the sea of low lying eigenvalues. Questions about the
scaling of $\Omega$
with $N$ and the corresponding characteristic height and distribution of the eigenvalues that lie farthest away from the real line are natural and interesting in this context. These are open questions in the mathematics and mathematical physics literature on the subject.

Apart from the mathematical curiosity, there is also motivation coming from physics. In the physics literature, the eigenvalues of $J$ are associated with the  {\em zeroes} of a scattering matrix in the complex energy plane, and their complex conjugates with the   {\em poles} of the same scattering matrix, known as ``resonances''.
The latter are obviously the eigenvalues of matrices \eqref{defens} -- \eqref{defens_Gamma} with $\gamma$ replaced by $ -\gamma$. In that context the absolute value of the eigenvalue's imaginary part is associated with the ``resonance width''. The eigenvalues close to real axis are called ``narrow resonances'' and the outlier is called the ``broad resonance''.
The use of the Gaussian Unitary Ensemble for $H$ is justified by invoking the so-called Bohigas-Giannoni-Schmidt
conjecture \cite{BGS} describing spectral statistics of highly excited energy levels of some classes of systems whose classical counterparts are chaotic. The resulting ensemble $J$ is then an important ingredient in characterising statistical properties of scattering matrices in systems with quantum chaos and no time-reversal invariance, see \cite{FyodSomm97} for description of the associated framework going back to the pioneering paper \cite{VWZ85} . In that framework, the phenomenon of the outlier separation and the simultaneous movement of the rest of the eigenvalues
 towards the real axis was first discussed, albeit at a heuristic level, already in early theoretical works \cite{SokZel89,trapping1}, the latter work even establishing the correct asymptotic position of the outlier.  Later on, this phenomenon got considerable attention under the name ``resonance trapping'' and eventually was observed in experiments \cite{trapping_exp}.

\begin{figure}[t]
\includegraphics[width=.48\linewidth]{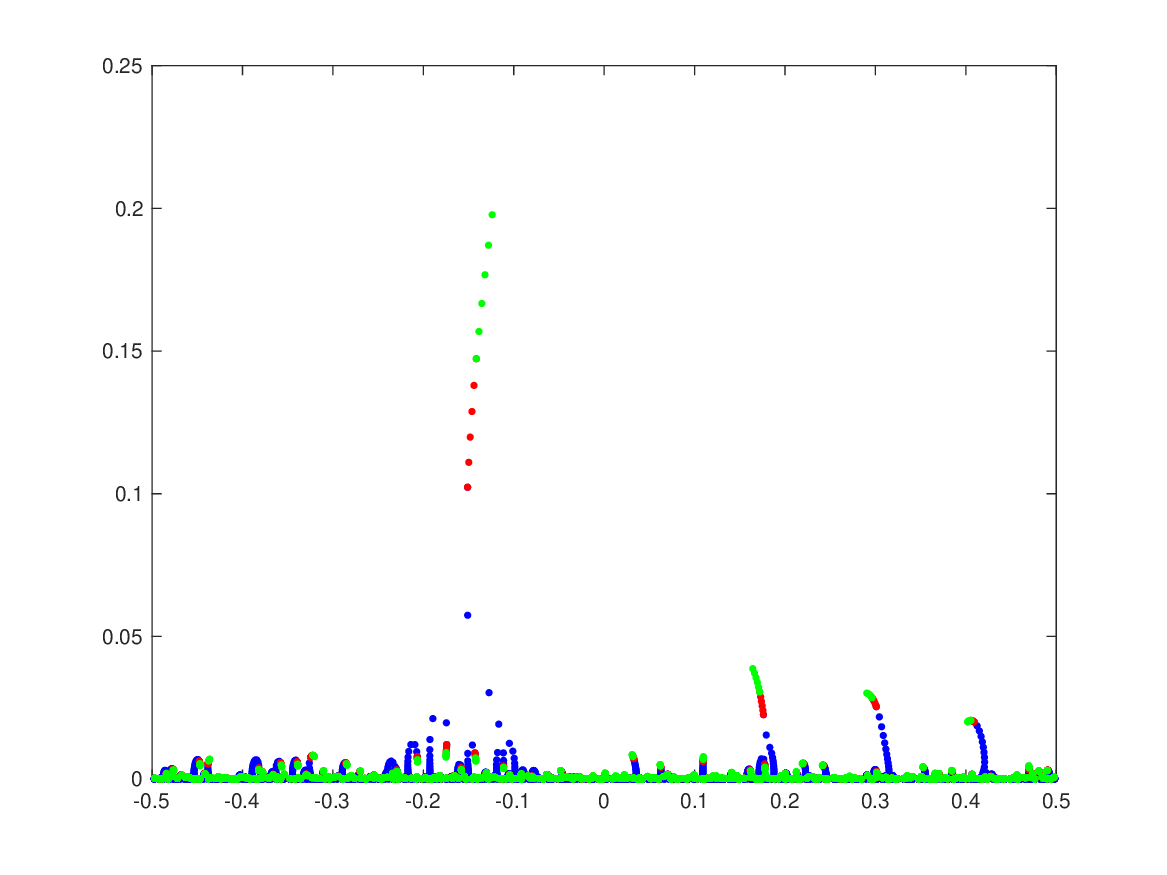}
\includegraphics[width=.48\linewidth]{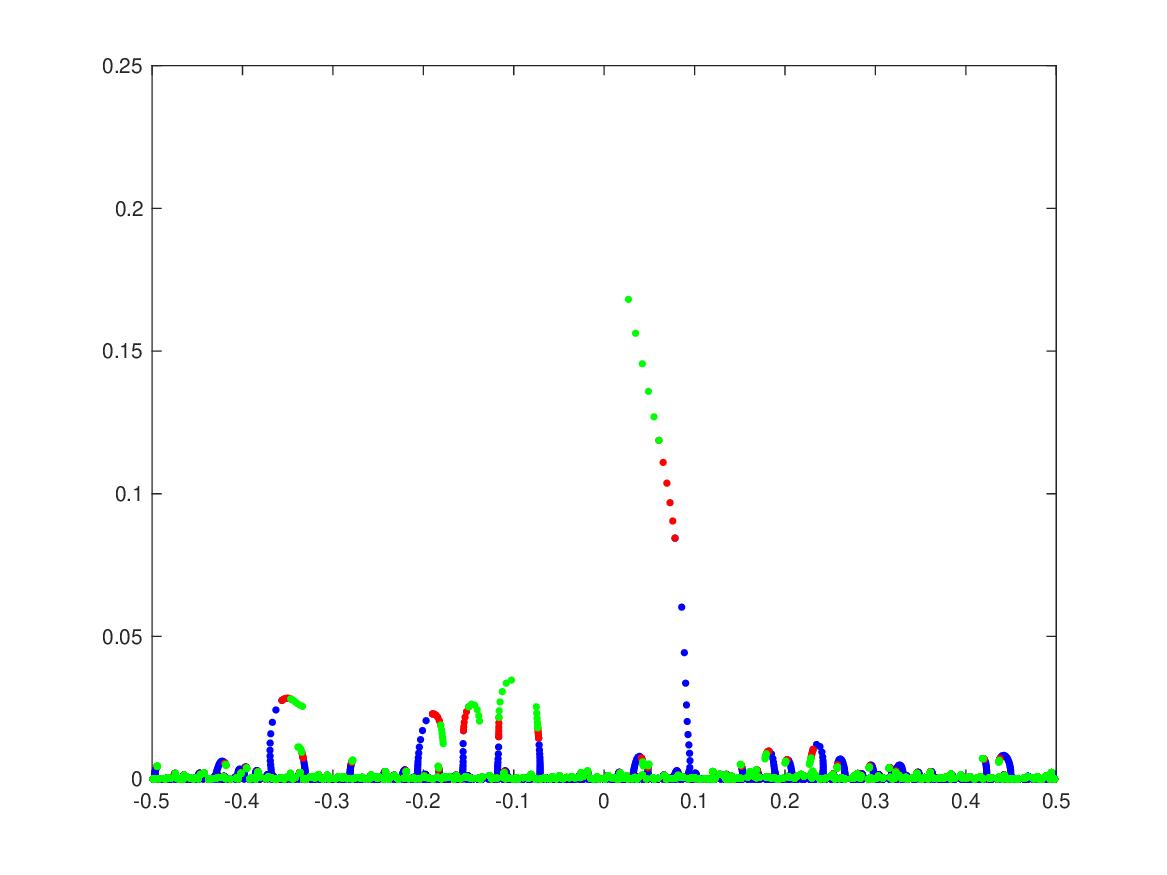}
\caption{$\gamma$- trajectories of eigenvalues of matrices  \eqref{defens} -- \eqref{GUE} of dimension $N=1000$ near the origin. Each plot represents a different sample of $H$ from the GUE \eqref{GUE}.
The parameter $\gamma$ is varying in the interval  $[0,0.5]$
in the increments of $0.05$ (blue dots), in the interval $[0.5, 1]$ in the increments of $0.1$ (red dots), and  in the interval $[1,1.5]$ in the increments of $0.1$ (green dots).
}
\label{Fig:1}
\end{figure}

Very recently, Dubach and Erd\H{o}s \cite{DubErd2021} performed a detailed analysis of the eigenvalue trajectories, with respect to changing the parameter $\gamma$, in the random matrix ensemble $H+i \gamma  v v^{*} $ in the settings when $H$ is assumed to be a Wigner matrix and $v$ a column vector of unit length.
It turned out that the evolution of the eigenvalues is governed by a system of \emph{deterministic} first-order differential equations subject to random initial conditions, with the initial positions and velocities expressed in terms of the eigenvalues and eigenvectors of $H$. In addition, 
under suitable conditions on the distribution of matrix entries of $H$ ensuring the validity of the uniform isotropic local law (Theorem 5 in \cite{DubErd2021} ), Dubach and Erd\H{o}s  proved
 that with high probability the eigenvalue outlier is distinctly separated from the rest of the eigenvalues for all
\begin{align}\label{eq:supercrs}
\gamma > 1+\frac{N^{\varepsilon}}{\sqrt[3]{N}},  \quad \varepsilon>0.
\end{align}
Moreover, if $\varepsilon <{1}/{3}$, i.e. if $N^{-1/3+\varepsilon}$ is asymptotically small, the outlier's height is $2N^{-1/3+\varepsilon}$ and its real part is in the window of width $N^{-1/3-\epsilon/4}$ around the origin, whereas all other eigenvalues are no higher than $N^{-1/3-\varepsilon}$. In addition, with high probability, for all
\begin{align}\label{eq:subcrs}
\gamma < 1-\frac{N^{\varepsilon}}{\sqrt[3]{N}},  \quad \varepsilon>0,
\end{align}
 no eigenvalue reaches the heights
\begin{align}\label{eq:chh}
Y=\frac{m}{\sqrt[3]{N}}, \quad m>0.
\end{align}
These findings suggest that the width $\Omega$ of the transition region around $\gamma=1$ 
 scales with as $N^{-1/3}$ for $N$ large. 
 Naturally, for $\gamma$ inside this region one would expect to find several eigenvalues, including the emerging ``atypical'' outlier, with imaginary parts on the critical scale \eqref{eq:chh} much exceeding the height $O(N^{-1})$ of low lying eigenvalues, as illustrated in Figure~\ref{Fig:1}.
One might call such eigenvalues "typical extremes" to emphasise atypicality of the emerging outlier.

To a large extent our paper is motivated by \cite{DubErd2021} and aims to provide 
quantitative insights into this picture of the outlier emerging from the cloud of extreme eigenvalues.
Whilst the approach of Dubach and Erd\H{o}s is dynamical (fix matrix $H$ and study eigenvalue trajectories as the magnitude $\gamma$ of the deformation increases), our approach is statistical (fix a scale for $\gamma$ and count the number of eigenvalues
on characterisitc spectral scales in the complex plane
averaged over the distribution of $H$ which, for technical reasons, we assume to be GUE). {{Our present approach is limited to the expected values; analysing higher order moments is left as an interesting problem for future investigations. However, even with such a basic tool we are able
to develop rather detailed quantitative understanding of the outlier separation and the associated restructuring transition in the spectra of matrices $J$}}.

As such, the two approaches complement each other very well. For example, we prove that for
\begin{align}\label{eq:crs}
\gamma=1+\frac{\alpha}{ \sqrt[3]{N}}, \quad \alpha \in \R,
\end{align}
the expected number of eigenvalues whose height exceeds the level \eqref{eq:chh} is asymptotically given by the integral $\int_m^{\infty} p^{(\im)}_{\alpha}(m^{\prime}) d m^{\prime} $ with density
\begin{align*}
\widetilde p^{(\im)}_{\alpha}\!(m) = \frac{1}{2\sqrt{\pi}}  \frac{\frac{3}{2m}+\left(\frac{3m}{2}-\alpha\right)^2}{m^{3/2}}\,
 e^{-m \left(\alpha-\frac{m}{2}\right)^2}, \quad m>0.
\end{align*}
This density is 
the average density  of the extreme eigenvalues at height \eqref{eq:chh}.
Together with findings in \cite{DubErd2021} this result establishes that the width $\Omega$ of the transition region around $\gamma=1$ indeed scales with $N^{-1/3}$. Similarly, we are able to determine the average density of extreme eigenvalues $Z_j=X_j+iY_j$ of $J$ near the origin in the complex plane 
in the critical scaling regime when when $q+im=\sqrt[3]{N} Z = O(1)$. As a function of coordinates $q$ and $m$, this density, when appropriately rescaled,
is given by
 \begin{align*}
\widetilde p_{\alpha} (q, m)=
 \frac{1}{4\pi m}\left[\frac{1}{m}+\frac{q^2}{ 4}+\left(\frac{3m}{2}-\alpha\right)^2\right]
 e^{-m\left[\!\frac{q^2}{4}+\left(\alpha-\frac{m}{2}\right)^2\right]}\, , \quad q\in \R, m>0.
\end{align*}
It can be verified that $\int_{-\infty}^{+\infty} \widetilde p_{\alpha} (q, m) dq =\widetilde p^{(\im)}_{\alpha}\!(m)$, implying that the population of extreme eigenvalues at the critical height \eqref{eq:chh} which generates the eventual outlier (as  $\alpha $ is approaching  infinity) is constrained to a narrow vertical strip of width $O(N^{-1/3})$ about the origin (the centre of the eigenvalue band of $H$). Thus our results
 both confirm and complement the analysis in \cite{DubErd2021}, and show that it indeed touched
   the optimal scales in $\gamma$ \eqref{eq:crs}, both along the real and imaginary axes.

We would like to conclude this section with a short description of the structure of our paper. In Section 2 we develop quantitative heuristic analysis of the outlier separation. This analysis elucidates the emerging critical scaling in $\gamma$ and the critical spectral scalings in the complex plane and provides a useful background for rigorous calculations later on. This section also offers our outlook on the universality of the exponent $-1/3$ in \eqref{eq:crs}. Section 3  contains the statement of our main results and discussion.
In Section 4 we express the expected density of eigenvalues of $J$ and the density of their imaginary parts at finite matrix dimensions in terms of, respectively, Hermite and Laguerre polynomials. These expressions are then used  in Sections 5 and 6 for asymptotic analysis of eigenvalue densities in various scaling limits. The two appendices contain derivations of technical auxiliary results.

\section{Low lying eigenvalues and their extremes: a heuristic outlook}
\label{Sec:2}

Before presenting our main results in the next Section, we would like to offer our quantitative heuristic insights into the outlier separation elucidating the emerging scalings and mechanisms behind them and providing a useful background for rigorous calculations later on.

With $z_j=X_j+iY_j$ standing for the eigenvalues of matrices $J=H+i\Gamma$, the angular brackets $\left\langle ... \right\rangle$ standing for averaging over the GUE matrix $H$ \eqref{GUE}, and $\delta(X)$ for the Dirac delta-function, the expected number of eigenvalues of $J$ in domain $D$ can be computed by integrating the mean eigenvalue density
\begin{align}\label{meandenjoint}
\rho_N(X,Y)=\Big\langle\frac{1}{N}\sum_{j=1}^N \delta(X-X_j) \delta(Y-Y_j)\Big\rangle
\end{align}
over $D$ and multiplying the result by $N$.
For example, the expected number  $\mathcal{N}_{\gamma}(Y)$ of the eigenvalues of $J$ which lie above the line $\im z =Y$ in the complex plane is given by the integral
\begin{align}\label{Ngamma}
\mathcal{N}_{\gamma}(Y)=N\!\int_{-\infty}^{\infty}\int_Y^{\infty} \!\!\rho_N(X,Y^{\prime})\,dXdY^{\prime} = N\!\int_Y^{\infty}\!\!\rho^{(\im)}_N(Y^{\prime})\,dY^{\prime}\, ,
\end{align}
where $\rho^{(\im)}_N\!(Y)$ is the mean density of the imaginary parts {\em irrespective of the value of the real part},
 \begin{align}\label{meandenim}
\rho^{(\im)}_N(Y)
=\Big\langle \frac{1}{N}\sum_{j=1}^N\delta(Y-Y_j)\Big\rangle.
\end{align}

 Guided by the eigenvalue perturbation theory one can expect that the typical height $Y$ of the eigenvalues whose real part is close to a point $X\in (-2,2)$ in the spectral bulk scales with the mean separation  $\Delta=(N\nu(X))^{-1}$ between neighbouring real eigenvalues of the GUE matrix $H$ in the limit $N\to \infty$. On a more formal level, introducing the scaled version of $\rho_N(X,Y)$
\cite{FyodSomm96}
\begin{align}\label{meandenscaled}
{\widetilde \rho}_N(X,y):=\frac{1}{\nu(X)}\Big\langle \frac{1}{N}\sum_{j=1}^N \delta\big(X-X_j\big) \delta\big(y-2\pi\nu (X) N Y_j\big)\Big\rangle, \quad  -2 \!< \!X \!< 2\, ,
\end{align}
one finds that such scaled density is well-defined in the limit of large matrix dimensions \cite{FyodSomm96,FyodSomm97,FyodKhor_systematic_1999,FyodSommRev03}: for every $y>0$
\begin{align}\label{limmeandenscaled}
\widetilde \rho(X,y):= \lim_{N\to \infty}{\widetilde \rho}_N(X,y) =
 \displaystyle{ -\frac{d}{dy}\left[ e^{-yg(X)}  \frac{\sinh{y}}{y} \right] },  \quad g(X)=\frac{\gamma+\frac{1}{\gamma}} {2\pi \nu(X)} \, ,
 \end{align}
confirming that locally the typical height of low lying eigenvalues scales with $\Delta=(N\nu(X))^{-1}$.

Globally, the typical height of low lying eigenvalues scales with $N^{-1}$. Intuitively, this can be understood from the exact sum rule
\begin{align}\label{sum_rule}
\sum_{j=1}^N Y_j =\gamma
\end{align}
which follows from the obvious relation $\Tr J= i \gamma+ \Tr H$.  On a more formal level, consider the expected fraction of the eigenvalues of $J$ which lie above the level $\im z =Y$, and set $y=NY $.
In the limit $N\to\infty$,
\begin{align}
\label{eq:add1}
\frac{1}{N}\,  \mathcal{N}_{\gamma}\left(\frac{y}{N}\right)  &\sim 
\int_{-2}^{2}dX \, \nu(X)  \int_{2\pi \nu(X)y}^{\infty} dy^{\prime}\,   {\widetilde \rho}(X,y^{\prime})
\\[1ex]
\label{eq:add2}
& = 
\frac{e^{-y\left(\gamma+\frac{1}{\gamma}\right)}}{y}\, \int_{-2}^2  \frac{dX}{4\pi}  \left(e^{y\sqrt{4-X^2}}-e^{-y\sqrt{4-X^2}}\right) \, .
\end{align}
The integral in \eqref{eq:add2}
is the modified Bessel function $I_1(2y)$. Therefore,
\begin{align}\label{101}
\lim_{N\to\infty}\frac{1}{N}\,  \mathcal{N}_{\gamma} \left(\frac{y}{N}\right) = \frac{e^{-y \left(\gamma+\frac{1}{\gamma}\right)}}{ y}I_1(2y)\, .
\end{align}
From this,
\begin{align} \label{rhotilde}
\widetilde \rho^{(\im)} (y)\!\!:= \!\!\lim_{N\to \infty}\frac{1}{N}\, \rho^{(\im z)}_N\!\left(\frac{y}{N}\right)
&=-\frac{d}{dy} \left[ \frac{e^{-y \left(\gamma+\frac{1}{\gamma}\right)}}{ y}I_1(2y)\right] \\
\label{denimy1}
&=\frac{e^{-y \left(\gamma+\frac{1}{\gamma}\right)}}{ y} \left[ \left(\gamma+\frac{1}{\gamma}-2\right)I_1(2y)- I_0(2y) -I_2(2y) \right] .
 \end{align}
The density
$\widetilde \rho^{(\im)} (y)$
is the mean density of the scaled imaginary parts $y_j=NY_j$ in the limit of large matrix dimensions. Even though it describes low lying eigenvalues it contains some useful information about eigenvalues higher up in the complex plane.

As an example,  consider  the expected value of the sum of the imaginary parts of low lying eigenvalues.
Using definition (\ref{meandenim}), the sum rule (\ref{sum_rule}) implies that
\begin{align}\label{meansumrule}
N\int_0^{\gamma} Y \rho^{(\im)}_N(Y) dY=\gamma\, .
\end{align}
Upon rescaling $y=NY$, 
one could naively jump to the conclusion that  $\int_0^{\infty}  y\,  \widetilde \rho^{(\im)} (y) \,  dy\,=\gamma$.
However, by making use of \eqref{rhotilde} and integral 6.623(3) in \cite{GR2007}, one actually finds that
\begin{align*}
\int_0^{\infty}  y\,  \widetilde \rho^{(\im)} (y) \,  dy\, = \int_0^{\infty}   \frac{e^{-y \left(\gamma+\frac{1}{\gamma}\right)}}{ y}I_1(2y)\, dy = \frac{ \gamma + \frac{1}{\gamma} - \sqrt{\left( \gamma + \frac{1}{\gamma} \right)^2 - 4}}{2} = \begin{cases}
\gamma, & \text{if $\gamma <1$,}\\[2ex]
\frac{1}{\gamma}, & \text{if $\gamma >1$.}
\end{cases}
\end{align*}
Thus, if $\gamma<1$ then the imaginary parts of low lying eigenvalues indeed add up to $\gamma$, in full agreement with the sum rule \eqref{meansumrule}, whereas if $\gamma >1$ they add up only to $\frac{1}{\gamma}<\gamma$. The sum rule deficit $\gamma - \frac{1}{\gamma}$ is exactly the imaginary part of the outlier, and suggests that the rescaled limiting density of low lying eigenvalues, $ \widetilde \rho^{(\im)} (y)$, precisely misses the delta-functional mass  $\frac{1}{N}\, \delta \left(y- \left(\gamma - \frac{1}{\gamma}\right)\right)$.

As another example, consider the asymptotic form of $\widetilde \rho^{(\im)} (y)$ when $y\gg 1$.
It is markedly different depending on whether $\gamma=1$ or not. In the later case, using  in \eqref{rhotilde} the asymptotic expansion for the modified Bessel function of large argument, $I_p(x) \sim  \frac{e^x}{\sqrt{2\pi x}} (1-\frac{4p^2-1}{8x}+\ldots ) $ one finds   an exponential decay, whilst in the former case the decay is algebraic:
\begin{align}  \label{rhotildetail1}
\widetilde \rho^{(\im)} (y) =
\begin{cases}
\displaystyle{
\frac{e^{-y\frac{(1-\gamma)^2}{\gamma}}}{2\sqrt{\pi}\, y^{3/2}}
}
\left[  \frac{(1-\gamma)^2}{\gamma}  + \frac{30-3(\gamma+\gamma^{-1})}{16 y} + O\left(\frac{1}{y^2}\right)\right]
 & \text{if $\gamma\not= 1$}, \\[3ex]
\displaystyle{
\frac{3}{4\sqrt{\pi}}\frac{1}{y^{5/2}} +O\left(\frac{1}{y^{7/2}}\right)
}
  & \text{if $\gamma= 1$}\, .
\end{cases}
\end{align}

\medskip

It is instructive to return to the unscaled imaginary part $Y$ and take a closer look at
 the expected number  of the eigenvalues of $J$ exceeding  the level $\im z= Y$ in the limit
 $N\to\infty$.
 It is evident from \eqref{101} that
 \begin{align}  \label{heuristic}
\mathcal{N}_{\gamma}(Y)\sim 
\frac{e^{-NY\left(\gamma+\frac{1}{\gamma}\right)}}{ Y}I_1(2NY)\, ,
\end{align}
provided $NY=O(1)$.
Extending
this asymptotic relation
to
large values of $NY$ allows one to get 
insights, even if only heuristically, about the
characteristic scale of the highest placed among the low lying eigenvalues.
Along these lines, we define the  \emph{characteristic scale of the height of typical extreme eigenvalues}  as such level $Y_{e}$ that the expected number of eigenvalues with imaginary part exceeding $Y_{e}$  is of order of unity:
 \begin{align}\label{extremes}
\mathcal{N}_{\gamma}(Y_e)=O(1).
 \end{align}
We add the word typical to exclude the atypical eigenvalue (the outlier) which is known to exist when  $\gamma>1$.  Now, assuming $NY_e$ to be large (but still anticipating $Y_e\ll 1$) one can replace the Bessel function in  \eqref{heuristic} by its corresponding asymptotic expression and approximate:
 \begin{align}  \label{heuristic_asy}
\mathcal{N}_{\gamma}(Y_e)\approx \frac{e^{-NY_e\frac{(1-\gamma)^2}{\gamma}}}{2\sqrt{\pi N}\, Y_e^{3/2}}, \quad \quad 1 \ll NY_e \ll N\, .
\end{align}
The condition in \eqref{extremes} then leads to two essentially different scenarios depending on the value of $\gamma$. Namely, for every fixed positive $\gamma\ne 1$ the characteristic scale of the typical extreme values is, to leading order in $N$,   $O(N^{-1} \ln{N} )$. On the other hand, if $\gamma=1$ then the typical extreme values
 raise  from the sea of low lying eigenvalues to a much higher height of $O (N^{-1/3})$.
 This change of scale for extreme values is easy to trace back to the emerging power-law decay  in the vicinity of $\gamma=1$ which is evident in \eqref{rhotildetail1}.

In fact, as evident from \eqref{heuristic_asy}, the typical extreme values scale as   $Y_e = O (N^{-1/3})$ not only at $\gamma=1$, but  also as long as $|1- \gamma|\propto N^{-1/3}$.
Actually, by setting simultaneously $\gamma=1+{\alpha}{N^{-1/3}}$ and $Y_e={m}{N^{-1/3}}$ the asymptotic relation (\ref{heuristic_asy}) is converted into
 \begin{align}\label{heuristic_asy_scaled}
\mathcal{N}_{1+\frac{\alpha}{\sqrt[3]{N}}}\left(\frac{m}{\sqrt[3]{N} } \right)\approx \frac{e^{-m\alpha^2}}{2\sqrt{\pi }\, m^{3/2}}\, ,
\end{align}
an expression that is indeed of order of unity for all fixed values of $\alpha$ and $m>0$.
Thus, the width of the transition region about $\gamma=1$ must scale with $N^{-1/3}$.
Combined with the existence of a distinct outlier at height $\gamma-\gamma^{-1}\gg Y_e$
one may indeed see that our heuristic argument perfectly agrees with the  conjecture of Dubach and Erd\H{o}s about the critical scaling $\gamma = 1+ O(N^{-1/3})$ where the separation of typical and atypical extreme values happens.

 \medskip

 Before continuing our exposition of the heuristics behind the restructuring of the density of complex eigenvalues we would like to make two remarks. 

\smallskip
\noindent {\bf Remark 1.} \quad
To make further contact with the standard subject of extreme value statistics, it is useful to recourse to
the classical theory of extreme values for i.i.d. sequences of random variables $y_1,\ldots, y_N$, a succinct albeit informal summary of which  can be found in, e.g., \cite{evs-rev}.
In that case the probability law of extreme values  is characterised by the tail behaviour of the ``parent'' probability density function (pdf) $p(y)$ of $y_j$ and is essentially universal in the limit $N\to \infty$. In our context, the pertinent case for comparison is that of non-negative continuous i.i.d. random variables with the parent distribution supported on the entire semi-axis $[0,\infty)$. Then only two possibilities may arise. Those sequences  which are characterised by the power-law decaying pdf $p(y)\sim Ay^{-(1+\alpha)}$, $\alpha>0$, as $y\to\infty$
have their extreme values scaling with  $(AN/\alpha)^{{1}/{\alpha}}$
and the distribution of their maximum, $y_{max}=\max(y_1,\ldots,y_N)$, after rescaling converges to the so-called Fr\'echet
law in the limit $N\to\infty$. In contrast, if the parent pdf decays faster than any power, e.g.,  if $\ln p(y)\sim -y^{\delta}$, $\delta>0$, then, to leading order, extreme values scale with $(\ln{N})^{{1}/{\delta}}$, and  the distribution of the largest value $y_{max}$, converges, after a shift and further rescaling, to the so-called Gumbel law. Although, the imaginary parts of complex eigenvalues in the random matrix ensemble \eqref{defens} -- \eqref{GUE}  are not at all independent (as is evident from their JPDF  \eqref{JPD} resulting in a non-trivial determinantal two-point and higher order correlation functions at the scale $N^{-1}$, see \cite{FyodKhor_systematic_1999}), our scaling predictions for the typical extreme eigenvalues are in formal correspondence with the i.i.d. picture: a Gumbel-like scaling (with $\delta=1$) if $\gamma\ne 1$ and a Fr\'echet-like scaling (with $\alpha=3/2$) if $\gamma=1$. This is exactly as would have been implied in the i.i.d. picture by the tail behaviour of the mean eigenvalue densities in the two cases in (\ref{rhotildetail1}).
This fact naturally suggests to conjecture Gumbel statistics for the typical largest imaginary part
 (excluding possible outlier) for any $\gamma\ne 1$,
 changing to a Fr\'echet-like law for $\gamma=1$, with a possible family of $\alpha-$ dependent nontrivial extreme value statistics in the crossover critical regime $\gamma=1+\alpha N^{-1/3}$.
 Although we are not able to shed light on the distribution of typical extreme eigenvalues in the random matrix ensemble \eqref{defens} -- \eqref{GUE}, we will discuss some results in that direction for a somewhat related model at the end of the next section.

\smallskip
\noindent {\bf Remark 2.} \quad
The phenomenon of resonance width restructuring  with increasing the coupling to continuum (controlled in the present model by the parameter $\gamma$) and the emergence of the broad resonance has many features in  common with the so-called super-radiant phenomena in optics. This is well known in the physics literature, see \cite{AZ2011} and references therein.  Here, we would like to point to a similarity of the spectral restructure in the random matrix ensemble \eqref{defens} -- \eqref{GUE}  to a process in a different physics context,
 the so-called ``condensation transition'' which occurs in models of mass transport when the globally conserved mass $M$ exceeds a critical value, see e.g. \cite{cond_trans_rev} for a review. In such a regime, the excess mass
forms a  localised in space condensate coexisting with a background fluid in which the remaining mass is evenly distributed over the
rest of the system. A particularly simple case for analysing the condensation phenomenon is when the
system has a stationary state such that probability of
observing a configuration of masses $m_i$ factorises into the form $\prod_i f(m_i)\delta(\sum_i m_i-M)$.
In that context again the tail behaviour of the ``parent'' mass density $f(m)$ plays important role.
Although we would like to stress again that in our model the imaginary parts of the complex eigenvalues are not independent, the analogy with the condensation phenomenon is quite evident.

\medskip

Essentially the same heuristic analysis as in the above helps to clarify the numerically observed fact of the outlier emerging mostly close to the origin of the spectrum $\re z =0$. From this angle it is instructive to ask what should be the scale of extreme values for eigenvalues satisfying $|\re z|<W$, that are sampled in a window of a small widths $ W\ll 1$ around the origin (still assuming typically many eigenvalues in the window, so that $W\gg \Delta\sim 1/N$).
The total mean number of eigenvalues in the window $W$ whose imaginary parts
exceed the level $Y$ (but still formally remain of the order of $1/N$)
is now given by
\begin{align}
\mathcal{N}_{\gamma,W}(Y)=\frac{e^{-NY\left(\gamma+\frac{1}{\gamma}\right)}}{4\pi Y}\left[T_W(NY)-T_W(-NY)\right], \quad T_W(NY)=2\int_0^{W}e^{NY\sqrt{4-X^2}}dX.
\end{align}
For $NY\gg 1$ the term  $T_W(-NY)$ is exponentially suppressed, while the integral in $T_W(NY)$ is dominated by $X\ll 1$ and with required accuracy yields the leading-order expression in the form:
\begin{align}
\mathcal{N}_{\gamma,W}(Y\gg 1/N)\approx \frac{e^{-NY\left(\frac{(1-\gamma)^2}{\gamma}\right)}}{2\pi Y^{3/2}}\sqrt{\frac{2}{N}}\int_0^{W\sqrt{NY/2}}e^{-\frac{t^2}{2}}dt.
\end{align}
Now, let us assume that both the width $W$ of the window and the parameter $\gamma$ scale with $N$ in this non-trivial way as
\begin{align}
W\sim N^{-1+\kappa}, \quad  0<\kappa\le 1, \quad \quad \text{and}\quad\quad \gamma=1-\alpha N^{-\delta}, \quad
  0<\delta\le \infty, \,\, \alpha\in \mathbb{R},
\end{align}
and again apply the same heuristic procedure to determine the scale of extreme values $Y_{e}(\kappa,\delta)$ in the window as $N\to \infty$ for given values of exponents $\kappa$ and $\delta$. A straightforward computation shows that the arising scale of extreme values very essentially depends on whether the parameter $\delta$ satisfies  $0<\delta<1/3$ or  $1/3\le \delta<1$.
In the former case we find
\begin{align}
Y_{e}(\kappa,0<\delta<1/3) \approx
\begin{cases}
\displaystyle{
 N^{-1+\kappa},
 }
  & \text{if }\,\,  0<\kappa<2\delta\, ,
  \\[1ex]
\displaystyle{
\frac{\kappa-2\delta}{\alpha}N^{-1+2\delta}\ln{N},
}  & \text{if }\,\, 2\delta<\kappa<1-\delta\, ,
\\[2ex]
\displaystyle{
 \frac{1-3\delta}{\alpha}N^{-1+2\delta}\ln{N},
 }
   & \text{if }\,\, 1-\delta<\kappa<1\, .
\end{cases}
\end{align}
whereas in the latter case
\begin{align}
Y_{e}(\kappa,1/3\le \delta<1) \approx
\begin{cases}
\displaystyle{
N^{-1+\kappa},
} & \mbox{if}\,\, 0<\kappa<2/3, \\[1ex]
\displaystyle{ N^{-1/3},
}  & \mbox{if}\,\, 2/3<\kappa<1.
\end{cases}
\end{align}
One may say that as long as  $\delta<1/3$ the system is not fully in the well-developed ``critical regime'', and the extreme value scale
is growing with the window width, saturating at the Gumbel-like scale $N^{-1+2\delta}\ln{N}$. At the same time, as long as $\delta$ exceeds the threshold value $\delta=1/3$, the typical extreme values reach the scale  $Y_e =O\left(N^{-1/3}\right)$ as long as
they are sampled in a window of width exceeding the scale $W_c=O\left(N^{-1/3}\right)$, thus
 containing $ O(N^{2/3})$ eigenvalues. This heuristics suggests that only eigenvalues satisfying
 $|X|<W_c$ typically have a nonvanishing probability to reach to the maximum height in the complex plane, and eventually to generate an outlier as  $\alpha$ increases. It would be also natural to expect the corresponding extreme eigenvalues to follow the Fr\'echet-type statistics for their imaginary parts, as opposed to the Gumbel statistics in the former case.

 \medskip

 We would like to end our heuristic considerations with a brief heuristic outlook on the universality of the scaling factor $N^{-1/3}$
which is key to the correct description of the transition in question.  As is evident from (\ref{heuristic_asy}) the exponent $-1/3$  is implied by the scaling law
\begin{align}\label{eq:scale}
\mathcal{N}_{\gamma=1}(Y) \propto  \frac{1}{N^{1/2} Y^{3/2}}
\end{align}
in the limit $NY \gg 1$ for the expected number of eigenvalues exceeding the level line $\im z =Y$. Thus, to investigate the extent of universality of this exponent one needs to trace the origin of the scaling law \eqref{eq:scale}. This can be readily done by returning to  the asymptotic relation \eqref{eq:add1}--\eqref{eq:add2} which was used to obtain \eqref{eq:scale}. On evaluating the integral in \eqref{eq:add2} for large values of $y=Y/N$ by the Laplace method it becomes immediately apparent that the power $Y^{-3/2}$ on the right-hand side in \eqref{eq:scale} and, hence, the exponent in question stems from the quadratic shape of the limiting GUE eigenvalue density function $\nu(X)=(2\pi)^{-1}\sqrt{4-X^2} $ in the vicinity of its maximum.  It is natural to conjecture that had one started from a random Hermitian matrix $H$
taken from the broad class of invariant ensembles characterised by joint probability density function $\propto\exp{-N\Tr V(H)}$ with a suitable potential $V(H)$ (or from the class of Wigner matrices with suitable conditions on the iid entries), the asymptotic expression (\ref{limmeandenscaled}) for the scaled eigenvalue density ${\widetilde \rho}_N(X,y)$ would retain its validity after replacing $\nu(X)$ in \eqref{meandenscaled}--\eqref{limmeandenscaled}  by the corresponding limiting eigenvalue density of $H$. For example, as was shown albeit not fully rigorously in \cite{FKS1998},  such universality of the scaled eigenvalue density near the real line is exhibited by almost Hermitian random matrices which are morally similar to finite rank non-Hermitian deviations  as in \eqref{defens}--\eqref{defens_Gamma}. 
Since asymptotic relation \eqref{eq:add1}--\eqref{eq:add2} is the immediate corollary of \eqref{limmeandenscaled}, one then concludes that as long as the limiting eigenvalue density of $H$ has a single
global parabolic-shaped maximum, an additive rank-one non-Hermitian deformation will demonstrate the same type of critical scaling for its extreme complex eigenvalues, and, most probably, after appropriate rescaling, the same type of critical behaviour of the density of imaginary parts as described in the next section. One can however imagine
invariant ensembles where the mean eigenvalue density would have a non-parabolic behaviour close to the maximum point.

From this point of view, the noticed in \cite{DubErd2021}  resemblance of the $N^{-1/3}$ critical scaling in the present model and the  edge scaling of extreme real eigenvalues of GUE, which, e.g., manifests itself in the so-called BBP \cite{BBP} transition under additive rank-one Hermitian perturbation of the GUE, looks to us purely coincidental. Indeed, the latter is known to have its origin in the square root behaviour of the mean density $\nu (X)$  at the spectral edges where $\nu(X)$ vanishes, and as such seems  to have nothing to do with the behaviour of the same density close to its maximal point.

\section{Main results and discussion}
\label{Sec:3}

Our first result concerns the mean density of imaginary parts $\rho^{(\im)}_N(Y)$ \eqref{meandenim} in the large deviation regime $Y\gg N^{-1}$.  We note that no eigenvalue of $J$ has imaginary part equal or greater than $\gamma$. This is a consequence of the sum rule \eqref{sum_rule}. Therefore we only consider the range of values $Y\in [0,\gamma)$.

\begin{theorem}\label{Thm_App 1}

Consider the random matrix ensemble  \eqref{defens} -- \eqref{GUE} in the scaling regime
\begin{align}\label{EValues}
N^{1-\epsilon}Y = y > 0, \quad 0<\epsilon\le 1, \quad N\to\infty \, .
\end{align}
Then for every fixed $\gamma >0 $ and $\epsilon\in (0,1]$
\begin{align}\label{LDform}
\rho^{(\im)}_N(Y)\sim \frac{1}{\sqrt{N}}\, \Psi_{\gamma}(Y)\, \exp{-N\Phi_{\gamma}(Y)},
\end{align}
with
\begin{align}\label{LDrate}
 \Phi_{\gamma}(Y) &=Y(\gamma-Y)-\ln \frac{\gamma-Y}{\gamma}-Yr_*(Y) +2\ln r_*(Y), \\[1ex]
\label{LDpreexp1}
 \Psi_{\gamma}(Y)&=\frac{1}{\sqrt{2\pi}}\frac{\gamma}{(\gamma-Y)^2}\frac{\left[1-r_*(Y)(\gamma-Y)\right]^2}{Y^{3/2}(Y^2+4)^{1/4}}\, ,
\end{align}
and
\begin{align}\label{rstar}
r_*(Y)=\frac{\sqrt{Y^2+4}-Y}{2}\, .
\end{align}
The rate function $\Phi_{\gamma}(Y)$ is a smooth non-negative function of $Y$ on the interval $[0,\gamma)$
vanishing at $Y=0$. The rate function is monotone increasing on this interval if $\gamma\le 1$, whereas if  $\gamma>1$ then it has two local extrema: a local minimum at $Y_*=\gamma-\gamma^{-1}$ where it vanishes, and a local maximum at $Y_{**}=\frac{2(\gamma-\gamma^{-1})}{3+\sqrt{1+8\,  \gamma^{-2}}} < Y_*$.

\end{theorem}

\medskip

By the way of discussion of the above Theorem a few remarks are in order.

\begin{figure}[b]
\includegraphics[width=.45\linewidth]{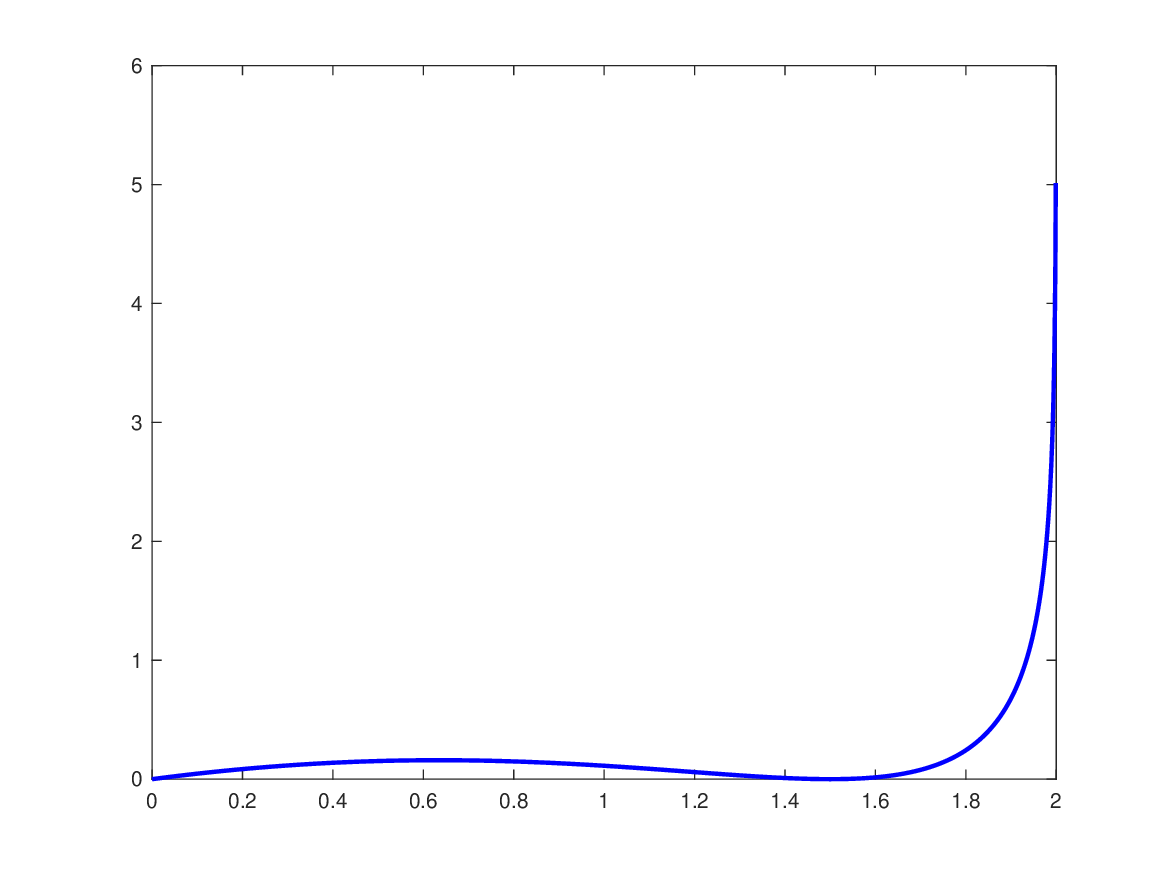}
\includegraphics[width=.45\linewidth]{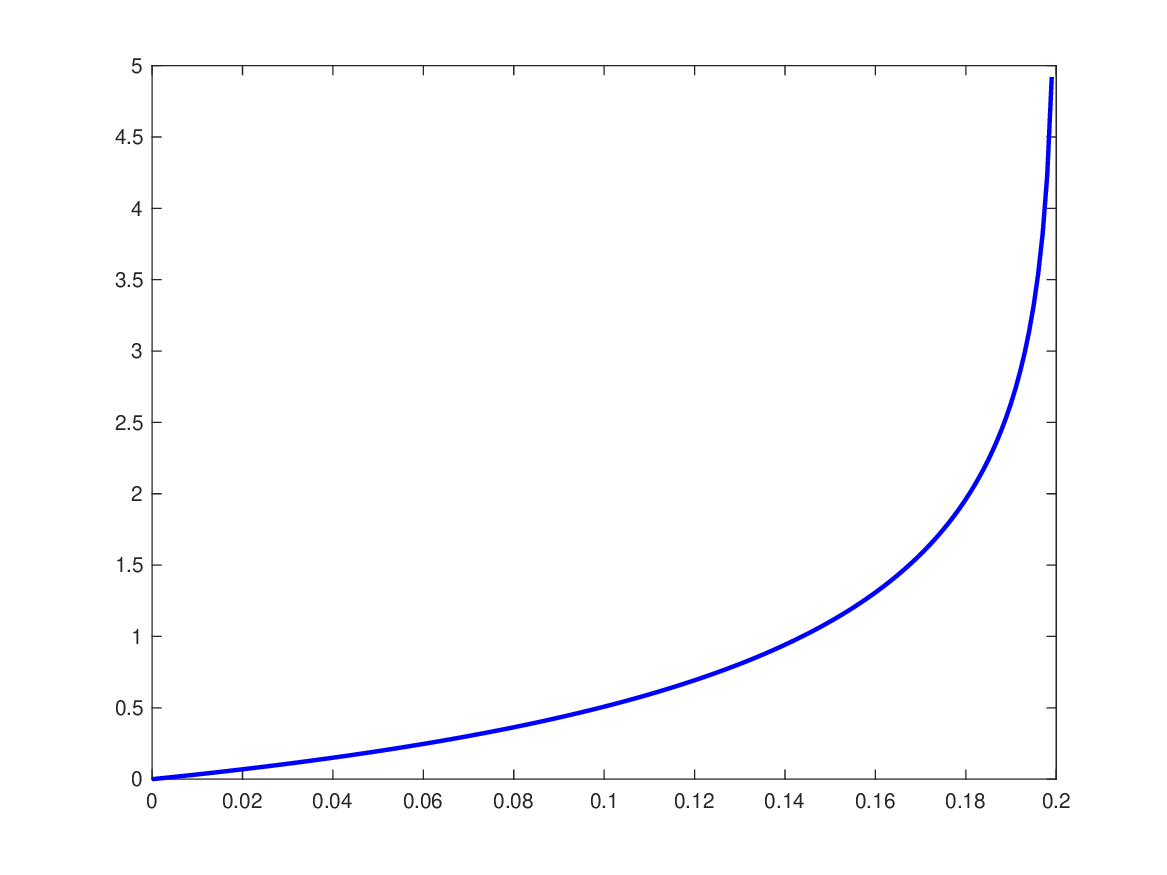}
\caption{\small Plots of the rate function $\Phi_{\gamma} (Y)$ for $\gamma=2$ (plot on the left) and $\gamma=0.2$ (plot on the right).}
\label{Fig:2}
\end{figure}

\smallskip
\noindent {\bf Remark 3.} \quad
The two distinct   profiles of the rate function are illustrated in Figure~\ref{Fig:2}.  If $\gamma >1$, the point $Y_*=\gamma-\gamma^{-1}$  where the Large Deviation Rate function $\Phi_{\gamma}(Y)$ vanishes can be identified as the most probable value of the imaginary part in the region $Y\gg N^{-1}$,
converging in the limit $N\to\infty$ to (the height of) the outlier, see next comment. At the same time, the other extremal point, $Y_{**}$, can be interpreted as  the \emph{true boundary}, along the imaginary axis in the complex plane, between the bulk of eigenvalues and the spectral outlier. This is because the pre-exponential factor $\Psi_{\gamma}(Y)$ in \eqref{LDform} vanishes at $Y=Y_{**}$ too. Hence,  $\rho^{(\im)}_N(Y_{**})\to 0$ in the scaling limit \eqref{EValues}.

\unskip

\begin{figure}[t]
\includegraphics[width=.45\linewidth]{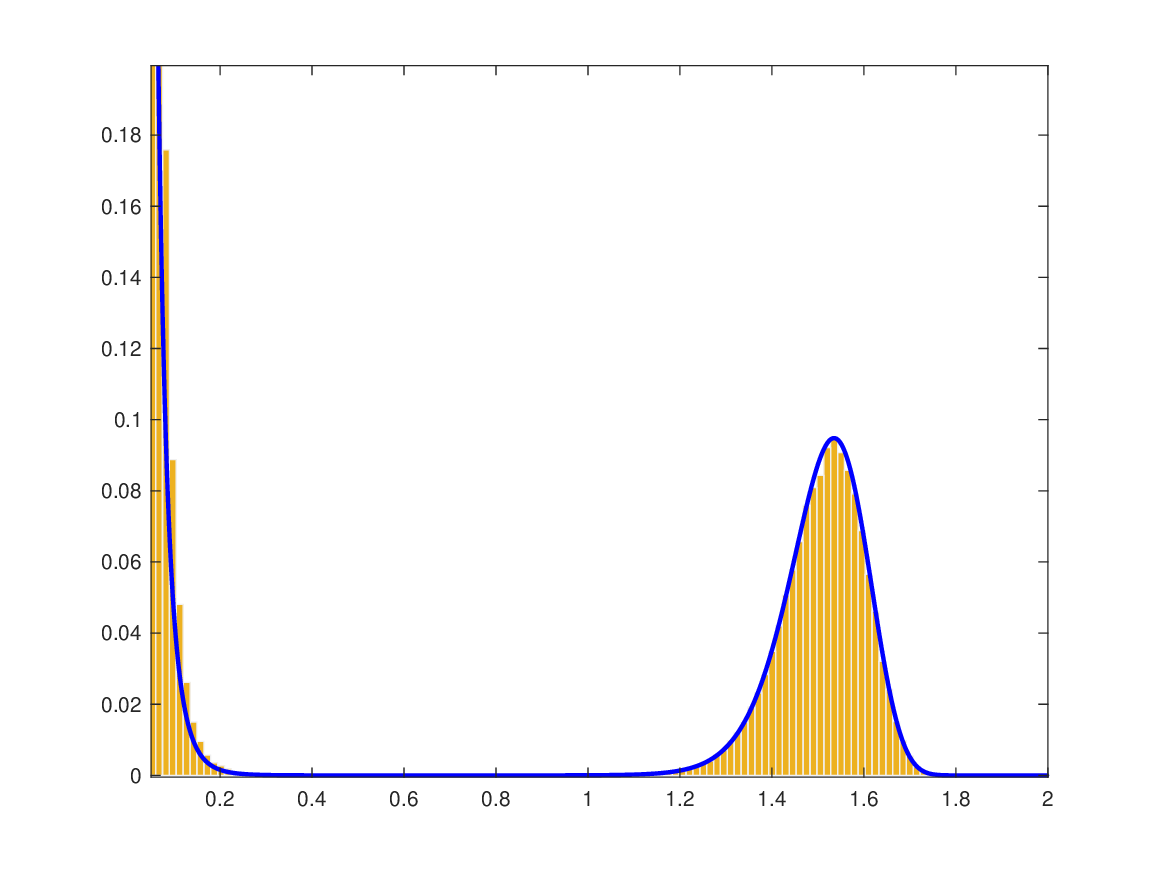}
\includegraphics[width=.45\linewidth]{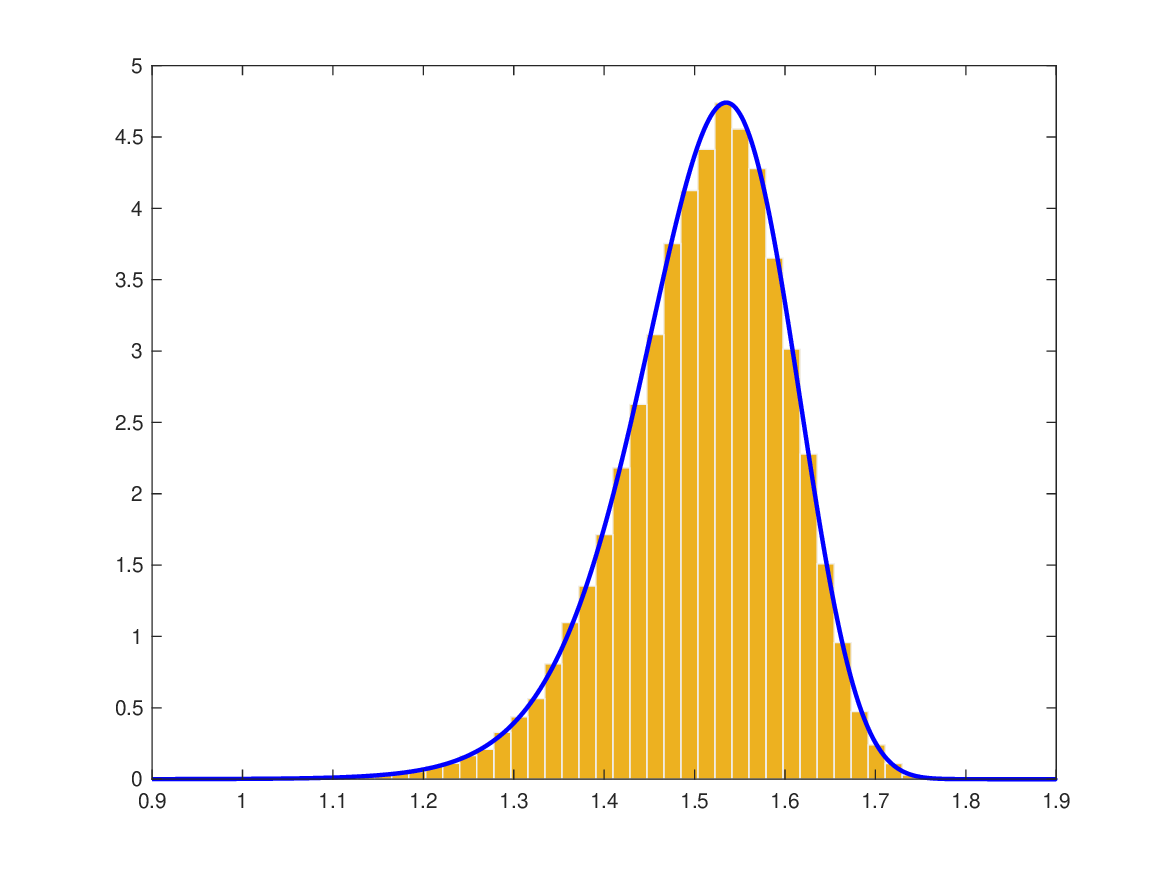}
\caption{\small Histograms of the imaginary parts $Y_j$ of the eigenvalues in the random matrix ensemble \eqref{defens} -- \eqref{GUE}. In both plots $N=50$ and $\gamma=2$. Plot on the left: Histogram of $Y_j$'s versus the large deviation approximation of density of the imaginary parts given by \eqref{LDform} (solid line). Plot on the right:  Histogram of the largest imaginary part $Y_{max}=\max Y_j $ versus the large deviation approximation $p_N(Y)$ \eqref{p_N}  of the p.d.f.  of $Y_{max}$ (solid line). Each plot was produced using 100,000 samples from the GUE distribution \eqref{GUE}.   }
\label{Fig:3}
\end{figure}
\unskip

\medskip

\smallskip
\noindent {\bf Remark 4.} \quad
The Large Deviation approximation \eqref{LDform} for $\gamma>1$ describes fluctuations of the imaginary part of the outlier around its most probable value $Y_*=\gamma-\gamma^{-1}$. The law of these fluctuations in the limit $N\to\infty$ can be easily determined from (\ref{LDform}). To this end, we first note that for $N$ large the magnitude of fluctuations about $Y_*$ scales with ${1}/{(\sqrt{N} |\Phi_{\gamma}^{\prime\prime}(Y_*)|)}$. Calculating the second derivative and rescaling the density $\rho^{(\im)}_N(Y)$ correspondingly, one finds (in the limit $N\to\infty$) that
\begin{align}\label{Gauform}
N\, \rho^{(\im)}_N\!\Big(Y_* + \frac{\sigma u}{\sqrt{N}}\Big)  \sim \frac{1}{\sqrt{2\pi }}e^{-\frac{u^2}{2}}, \quad \sigma^2=\frac{1}{\gamma^2}\frac{\gamma^2+1}{\gamma^2-1}\, .
\end{align}
The integral of the rescaled density on the left-hand side over the entire range of values of $u$ counts the expected number of eigenvalues in the $\frac{\sigma}{\sqrt{N}}$-neighbourhood of $Y_*$. Evidently,  this integral is approaching unity as $N\to\infty$, confirming that the rescaled density on the left-hand side in \eqref{Gauform} describes the law of fluctuations of a single eigenvalue - the outlier. Thus we recover one of the results of \cite{outlier} where laws of outlier fluctuations were established in greater generality than our assumptions  \eqref{defens_Gamma} -- \eqref{GUE}. We note that for finite but large values of $N$ the function
\begin{align}\label{p_N}
p_N(Y):=\sqrt{N}\, \Psi_{\gamma}(Y)\, \exp{-N\Phi_{\gamma}(Y)}
\end{align}
provides an approximation of the probability density function of the outlier $Y_{max}=\max Y_j$ in the interval $0< \varepsilon <Y <  \gamma$, $\gamma >1$.

In Figure~\ref{Fig:3} we plot histograms of the imaginary parts $Y_j$ of the eigenvalues and of their maximal value $Y_{max}=\max Y_j$  in the random matrix ensemble \eqref{defens} -- \eqref{GUE}  and make comparison with the corresponding Large Deviation approximations. Although the value of $N=50$ is only moderately large, one can observe a good agreement. Also, one can observe that the large-$N$  approximation \eqref{p_N} of the probability density of $Y_{max}$ captures well the skewness of the distribution of $Y_{max}$ for finite matrix dimensions. This skewness disappears in the limit $N\to \infty$, see equation \eqref{Gauform}.

\smallskip
\noindent {\bf Remark 5.} \quad
Consider now the scales $Y=O\left(N^{-1+\varepsilon}\right)$ with $\varepsilon \in (0,1)$. The expected number of eigenvalues with $N^{1-\varepsilon}Y \in [y_1,y_2] $ is given
by the integral
\begin{align}\label{expnum}
N \int_{y_1}^{y_2} \frac{1}{N^{1-\varepsilon}}\, \rho^{(\im)}_N\! \left( \frac{y}{N^{1-\varepsilon}}\right) dy.
\end{align}
The rescaled density in this integral can be found from \eqref{LDform} -- \eqref{LDpreexp1}:  
\begin{align}\label{eps}
\frac{1}{N^{1-\varepsilon}}\, \rho^{(\im)}_N\!\left( \frac{y}{N^{1-\varepsilon}}\right) \sim  \frac{1}{N^{\varepsilon/2}}\,\frac{1}{2\sqrt{\pi}} \frac{(1-\gamma)^2}{\gamma}  \frac{1}{y^{3/2}}  e^{-N^{\varepsilon}y\frac{(1-\gamma)^2}{\gamma }} \, , \quad \varepsilon \in (0,1)\, .
\end{align}
Evidently, if $\gamma\not=1$ then, away from the boundary point $y=0$, the integral in \eqref{expnum}  vanishes in the limit $N\to\infty$. Therefore for every fixed $\gamma\not=1$  and $0<\varepsilon<1$ there are no eigenvalues of $J$ whose imaginary part is scaling with $N^{-1+\varepsilon}$. On the other hand, according to the heuristics of Section \ref{Sec:2}, one should expect finite numbers of eigenvalues whose imaginary part is scaling with $N^{-1}\ln N$. These would be the extremes of the eigenvalues with the typical imaginary part  $Y=O(N^{-1})$.

By formally letting $\varepsilon \to 0$ in \eqref{eps} one obtains
\begin{align*}
\frac{1}{N}\, \rho^{(\im)}_N\!\left( \frac{y}{N}\right) \sim  \frac{1}{2\sqrt{\pi}} \frac{(1-\gamma)^2}{\gamma}  \frac{1}{y^{3/2}} \, e^{-y\frac{(1-\gamma)^2}{\gamma }} \, .
\end{align*}
This relation reproduces the leading order of the asymptotic form of the  density of the rescaled imaginary parts $y=NY$ in the region  $y\gg 1$, see the top line in \eqref{rhotildetail1}.  Thus, \emph{for a fixed value of} $\gamma\not= 1$ Theorem \ref{Thm_App 1} describes a crossover of the density of imaginary parts from  the characteristic scale of low lying eigenvalues to larger scales, including $Y=O(1)$ which is the scale of the outlier.

\medskip

Whereas the picture described by Theorem \ref{Thm_App 1} is quite complete for a fixed $\gamma$, it is not detailed enough to accurately describe 
the typical extreme eigenvalues  in the situation when the parameter $\gamma$ approaches its critical value $\gamma=1$ as $N$ is approaching infinity.
For example, from the heuristics of Section \ref{Sec:2} we know that both the width of the transition region about $\gamma=1$ and the height of the typical extreme eigenvalues scale with $ N^{-1/3}$. The Large Deviation approximation (\ref{LDform}), if applied formally in the transition region parametrised by $\gamma=1+\alpha N^{-1/3}$, yields  the following approximate expression for the rescaled density of imaginary parts:
\begin{align}\label{tr1}
\frac{1}{N^{1/3}} \, \rho^{(\im)}_N\!\left(\frac{m}{N^{1/3}}\right) \approx \frac{1}{N}  \frac{1}{2\sqrt{\pi}}
\frac{\left(\frac{3m}{2}-\alpha\right)^2}{ m^{3/2}} e^{-m (\alpha-\frac{m}{2})^2}\, .
\end{align}
Evidently, in the limit of small values of $m$ which corresponds to approaching the scale $Y=O(N^{-1})$ from above, this expression does not reproduce the correct power $5/2$ of algebraic decay \eqref{rhotildetail1} characteristic of this scale when $\gamma=1$. In contrast, the heuristics based on \eqref{heuristic}, see the approximations in \eqref{heuristic_asy} and  \eqref{heuristic_asy_scaled}, do reproduce the correct power. Indeed, by taking the derivative in $m$  of the expression on the right-hand side in \eqref{heuristic_asy_scaled}, one gets
\begin{align}\label{tr2}
\frac{1}{N^{1/3}} \, \rho^{(\im)}_N\!\left(\frac{m}{N^{1/3}}\right) \approx \frac{1}{N}  \frac{1}{2\sqrt{\pi}}
\frac{\frac{3}{2m}+\alpha^2}{ m^{3/2}} e^{-m \alpha^2 }\, .
\end{align}
In the limit of small values of $m$ the expression on the right-hand side agrees with the bottom line in  \eqref{rhotildetail1}.
One can also arrive at \eqref{tr2}  by making the formal substitution $\gamma=1+\frac{\alpha}{N^{1/3}}$ and $y=NY=mN^{2/3}$ in \eqref{rhotildetail1}.

\medskip

Our next Theorem is a refinement of Theorem \ref{Thm_App 1} in that it provides an accurate description of the density of the typical extreme eigenvalues in the transition region between the sea of low lying eigenvalues and the eigenvalue outlier.

\begin{theorem}\label{Thm_App 1a}
Consider the random matrix ensemble  \eqref{defens} -- \eqref{GUE} in the scaling regime
\begin{align}\label{sr}
\gamma=1+\frac{\alpha}{N^{1/3}}, \, Y=\frac{m}{N^{1/3}}, \quad N\to\infty\, .
\end{align}
Then, for every fixed  $\alpha\in \mathbb{R}$ and $m>0$,
 \begin{align}\label{scaleform}
 \frac{1}{N^{1/3}}\, \rho^{(\im)}_N\! \left( \frac{m}{N^{1/3}} \right) \sim  \frac{1}{N} \frac{1}{2\sqrt{\pi}}\frac{\frac{3}{2m}+\left(\frac{3m}{2}-\alpha\right)^2}{m^{3/2}}\,
 e^{-m\left(\alpha-\frac{m}{2}\right)^2} \, .
\end{align}
\end{theorem}

\medskip

This theorem confirms that the characteristic scale of the height of the typical extreme eigenvalues of matrix $J$ is $O\!\left(N^{-1/3}\right)$. Indeed,
the expected number
of  eigenvalues with imaginary part exceeding the level $Y=\frac{m}{N^{1/3}}$
is given by
\begin{align*}
N \int_m^{\infty}  \frac{1}{N^{1/3}} \, \rho^{(\im)}_N\!\left( \frac{m}{N^{1/3}} \right) \,dm\, ,
\end{align*}
which is a finite number in the limit $N\to\infty$.

\smallskip

 Theorem \ref{Thm_App 1a} also describes the density $\rho_N^{(\im)}(Y)$  in the cross-over
from the characteristic scale of low lying eigenvalues to the Large Deviation regime of Theorem \ref{Thm_App 1}. Indeed, for small values of $m$ the asymptotic expression \eqref{scaleform} matches the one in \eqref{tr2}, whilst in the limit of large values of $m$ it matches \eqref{tr1}.

\smallskip

The emerging outlier is captured by \eqref{scaleform} when \emph{both} $m$ and $\alpha>0$ are large. Intuitively this is clear from the comparison of (\ref{scaleform}) and \eqref{tr1}. On a more formal level,
one can come to the same conclusion by analysing the limiting density of extreme values
 \begin{align}\label{scaleform1}
 \widetilde p_{\alpha}^{(\im)}\!(m) = \frac{1}{2\sqrt{\pi}}\frac{\frac{3}{2m}+\left(\frac{3m}{2}-\alpha\right)^2}{m^{3/2}}\,
 e^{-m\left(\alpha-\frac{m}{2}\right)^2}, \quad m>0\, .
\end{align}
Using Wolfram Mathematica one finds
\begin{align*}
\frac{d}{dm}\, \widetilde p_{\alpha}^{(\im)}\!(m)= \frac{e^{-m\left(\alpha-\frac{m}{2}\right)^2}}{32\sqrt{\pi} \, m^{7/2}} \, Q_6(\alpha, m)\, ,
\end{align*}
where 
\begin{align*}
Q_6(\alpha, m)= -60 -48 \alpha^2m + 72 \alpha m^2 -16 \alpha^4 m^2 +80 \alpha^3 m^3 - 144 \alpha^2 m^4 +108 \alpha m^5 -27 m^6.
\end{align*}
Evidently, $Q_6(\alpha, m)<0$ for all $m>0$ if $\alpha$ is negative. Therefore, if $\alpha<0$ (subcritical values of $\gamma$) then the limiting density $ \widetilde p_{\alpha}^{(\im)}\!(m) $ is a monotonically decreasing function of $m$ on the entire interval $m>0$. One can interpret this profile as a population of extreme eigenvalues without an obvious ``leader''.  By continuity, this profile persevere for small positive $\alpha$. Indeed,  at $\alpha=0$ the polynomial $Q_6(0, m)$
has three pairs of complex conjugated roots, none are real. Since the roots of polynomials depend continuously on its coefficients, there exists an $\alpha_0>0$ such that for all $\alpha \in [0,\alpha_0]$ the polynomial  $Q_6(\alpha, m)$ in $m$ will still have no real roots and, hence, will take only negative values, implying that $ \widetilde p_{\alpha}^{(\im)}\!(m) $ is a monotonically decreasing function of $m$. By computing the roots of $Q_6(\alpha, m)$ in variable $m$, we can show that $0.6485 < \alpha_0 < 0.649$.

Once $\alpha > \alpha_0$, the polynomial $Q_6(\alpha, m)$ in $m$ acquires real roots.  In the limit of large positive $\alpha$ there are two real roots: to leading order these are
\begin{align*}
m_1 = 2\alpha \left(1+  \frac{3}{8\alpha^3} +o\left(\frac{1}{\alpha^3}\right) \right)  \quad \text{and} \quad
m_2 = \frac{2}{3} \alpha\left(1+  \frac{15}{8\alpha^3} +o\left(\frac{1}{\alpha^3}\right) \right)  .
\end{align*}
The larger root, $m_1$, is the point of local maximum of $\widetilde p_{\alpha}^{(\im)}\!(m)$, where $\widetilde p_{\alpha}^{(\im)}\!(m_1) \propto \alpha^{1/2} \gg 1 $, and the smaller root, $m_2$, is the point of local minimum $\widetilde p_{\alpha}^{(\im)}\!(m)$, where  $\widetilde p_{\alpha}^{(\im)}\!(m_2) \propto \alpha^{-5/2} \ll 1$. In fact, in the limit $\alpha \to\infty$ the larger root is transitioning into $Y_*$, the most probable value of imaginary parts, and, hence, it can be interpreted as the emerging spectral outlier. At the same time,
the smaller root  is transitioning into the true boundary $Y_{**}$ between the sea of low lying eigenvalues and the outlier. This cross-over can be validated by noticing that in the scaling limit \eqref{sr}  $Y_*=\gamma-\gamma^{-1} \sim 2\alpha$ and $Y_{**}= \frac{2(\gamma-\gamma^{-1})}{3+\sqrt{1+8\gamma^{-2}}} \sim \frac{2}{3}\alpha$.

\medskip

\begin{figure}
\includegraphics[width=.5\linewidth]{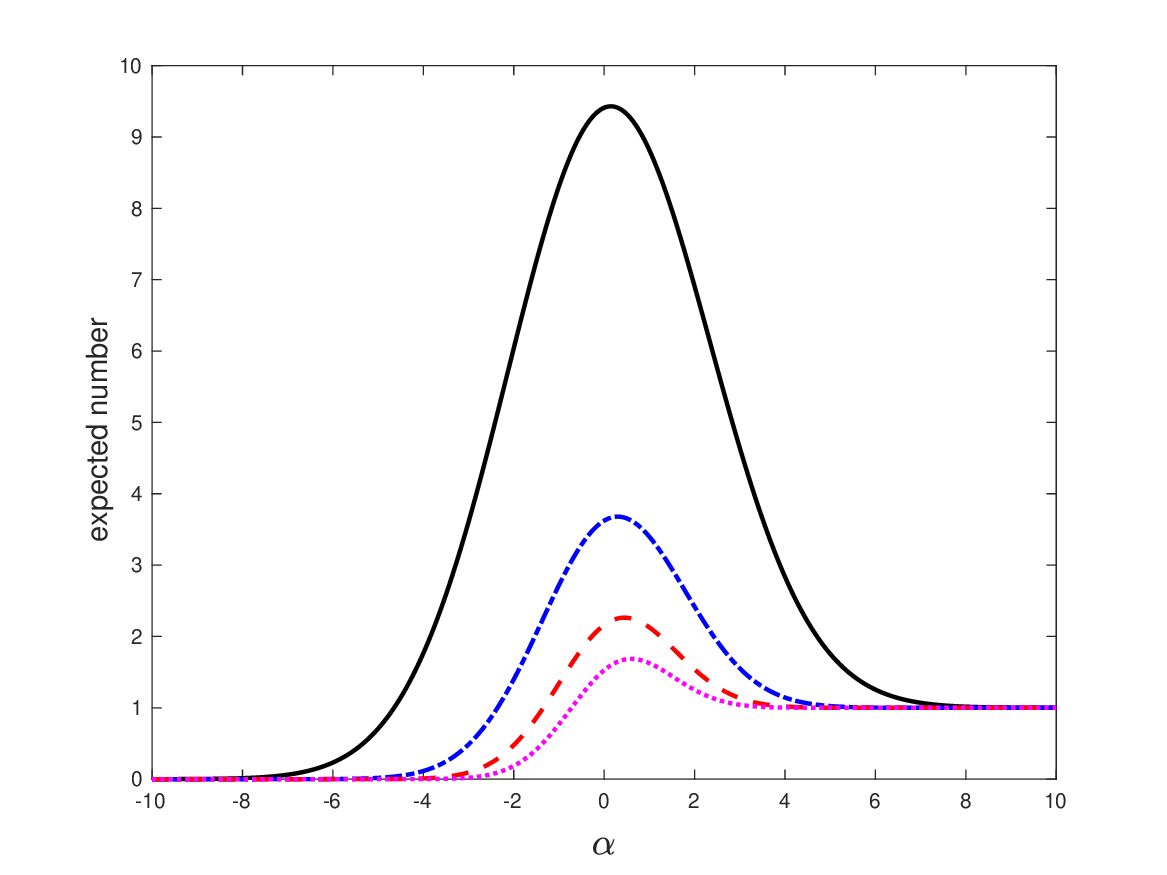}
\caption{\small Plot of the expected number $\widetilde{ \mathcal{N}}_{\alpha} (m) $ of the eigenvalues of $J$ with imaginary parts exceeding the level $ Y={m}{N^{-1/3}} $ as function of $\alpha$ when  $m=0.1$ (black solid line),  $m=0.2$  (blue dashdotted line), $m=0.3$ (red dashed line), and $m=0.4$  (magenta dotted line)}
\label{Fig:4}
\end{figure}

Further insights into the restructuring of the spectrum of $J$ can be obtained by looking at the $\gamma$-dependence of the expected number of the eigenvalues of $J$ with imaginary parts exceeding the level $ Y={m}{N^{-1/3}} $. In the scaling limit \eqref{sr} this number converges to
\begin{align*}
\widetilde{ \mathcal{N}}_{\alpha} (m) =  \int_m^{\infty} \widetilde p_{\alpha}^{(\im)}\!(m^{\prime}) dm^{\prime} \, .
\end{align*}
In Figure~\ref{Fig:4} we plot $\widetilde{ \mathcal{N}}_{\alpha} (m)$ as function of $\alpha$ for several values of $m$. One can observe that for any fixed $m>0$ the population of the extreme eigenvalues of $J$ that exceed the level  $Y={m}{N^{-1/3}} $   is, on average, growing as $\gamma$ is approaching the critical value $\gamma=1$ from below.  For $\gamma$  on the other side of $\gamma=1$, this population peaks a some point and then it starts to decline as $\gamma$ increases further, to a single eigenvalue which is the outlier.  All the other extreme eigenvalues are getting closer and closer to the real line with the increase of $\gamma$. One can think of them as being trapped in the sea of low lying eigenvalues. This picture is consistent with the eigenvalue trajectories of Figure~\ref{Fig:1} and provides a more quantitative description of the "resonance trapping" phenomenon \cite{trapping_exp} in the framework of random matrix theory.



Our final result aims to clarify the length of the central part of the spectrum of $J$ supporting nontrivial scaling behaviour of 
the extreme eigenvalues in the vicinity of the separation transition. To this end,
let us consider eigenvalues $z_j=X_j+iY_j$ of $J$ in the scaling regime when
\begin{align}\label{sr3}
\gamma=1+\frac{\alpha}{N^{1/3}}, \,\,  X=\frac{q}{N^{1/3}}\,\, , Y=\frac{m}{N^{1/3}}, \quad N\to\infty\, .
\end{align}
On average, eigenvalue numbers in this regime can be counted using the rescaled density
\begin{align*}
\widetilde p_{N} (q, m) :=
\left\langle  \sum_{i=1}^N\delta\left(q-N^{1/3}X_j\right) \, \delta\left(m-N^{1/3}Y_j\right)\right\rangle
=\frac{N}{N^{2/3}}\, \rho_N \left( \frac{q}{N^{1/3}}, \frac{m}{N^{1/3}} \right),
\end{align*}
where, as before, the angle brackets stand for the averaging over the GUE matrix $H$ in \eqref{defens} and $ \rho_N (X,Y) $ is the mean eigenvalue density  \eqref{meandenjoint}.

\medskip

 \begin{theorem}\label{Thm_App 1b}  Consider the random matrix ensemble  \eqref{defens} -- \eqref{GUE} in the scaling regime  \eqref{sr3}.
Then, for every fixed  $\alpha\in \mathbb{R}$, $q\in \mathbb{R}$ and $m>0$,
 \begin{align}\label{scaleformcent}
\widetilde p_{\alpha} (q, m):= \lim_{N\to\infty} \widetilde p_{N} (q, m)  =
 \frac{1}{4\pi m}\left[\frac{1}{m}+\frac{q^2}{ 4}+\left(\frac{3m}{2}-\alpha\right)^2\right]
 e^{-m\left[\!\frac{q^2}{4}+\left(\alpha-\frac{m}{2}\right)^2\right]}\, .
\end{align}
\end{theorem}

\medskip

It is easy to see from (\ref{scaleformcent}) that $ \int_{-\infty}^{\infty} \widetilde p (q, m) dq=\widetilde p^{(\im)}_{\alpha} (m)$.  Thus, Theorem \ref{Thm_App 1b} confirms the heuristics of Section \ref{Sec:2} in that the population of extreme eigenvalues which generates the eventual outlier (as  $\alpha $ is approaching  infinity) is constrained to a narrow vertical strip of width $O(N^{-1/3})$ about the origin.

\medskip

Our results demonstrate that despite being one of the simplest tools available, the mean eigenvalue density captures the eigenvalue and parameter scales associated with the spectral restructuring in the random matrix ensemble  \eqref{defens} -- \eqref{GUE}.
 However, it gives no information about finer details, such as  the probability distribution of the extreme eigenvalues during the restructure.  Calculating all the higher order eigenvalue correlation functions in the scaling regime \eqref{sr3} would be a significant step towards describing such finer details. Unfortunately, the eigenvalue point process in the random matrix ensemble  \eqref{defens} -- \eqref{GUE}  is not determinantal at finite matrix dimensions and such a calculation is a considerably more difficult analytic task compared to the mean eigenvalue density.

At this point we want to mention that the probability distribution of extreme eigenvalues can be determined in a related but different random matrix ensemble exhibiting a spectral restructuring not unlike one in \eqref{defens} -- \eqref{GUE}. This ensemble consists of subunitary matrices of the form
\begin{align}\label{subU}
J_{CUE}= U \diag (\sqrt{1-T}, 1, \dots, 1 )\, ,
\end{align}
where the matrix $U$ is taken from the Circular Unitary Ensemble (CUE) of complex unitary matrices uniformly distributed over $U(N)$ with the Haar's measure and $T\in [0,1]$ is a parameter. The ensemble was originally introduced in \cite{F00} and various statistical aspects of their spectra and eigenvectors were addressed in  \cite{FyodSommRev03,FM02,FK2007,KK2017} and most recently in \cite{FI2019}.

Obviously, if $T=0$ then the matrix $J_{CUE}$ is unitary and all of its eigenvalues  lie on the unit circle $|z|=1$. If $T>0$ and is fixed in the limit $N\gg 1$ then, typically, the eigenvalues of $J_{CUE}$ lie at a distance $O(N^{-1})$ from the unit circle with the farthest away being at a distance
$O\big(\frac{\log N}{(1-T)N}\big)$ with probability close to one. On the other hand, for $T=1$ one of the eigenvalues becomes identically zero, and the rest are distributed inside the unit circle in the same way as eigenvalues of the  so-called "truncated" CUE \cite{ZS2000}.

The similarity between the random matrix ensembles \eqref{subU} and \eqref{defens} -- \eqref{GUE} can be exemplified by analysing the mean density of the eigenvalue moduli $r_j=|z_j|$
\begin{align*}
\rho_N(r) = \Big\langle \frac{1}{N} \sum_{j=1}^N \delta (r-r_j)\Big\rangle_{CUE}
\end{align*}
in the limit of large matrix dimensions $N\to \infty$. One finds \cite{F00} that for
every fixed $T\in[0,1]$
\begin{align*}
\lim_{N\to\infty}N\rho_N(r) =
\begin{cases}
\delta(r) + \frac{2r}{(1-r^2)^2}, & \text{if $T=1$}, \\
0, & \text{if $0<T<1$}\, ,
\end{cases}
\end{align*}
whereas, on rescaling the radial density near the unit circle \cite{ZS2000,FyodSommRev03},  
\begin{align}\label{rhoU}
\widetilde \rho_{CUE} (y):=\lim_{N\to\infty}\frac{1}{N} \rho_N\left( 1-\frac{y}{N} \right) = - \frac{d}{dy} \left[ e^{-g y } \frac{\sinh y }{y}\right] \, ,\quad \text{with $g=\frac{2}{T}-1$.}
\end{align}
Equation \eqref{rhoU}
is identical, with the obvious correspondence
\begin{align}\label{gamma2T}
\frac{1}{2}\!\left(\gamma+\frac{1}{\gamma} \right) =\frac{2}{T}-1\, ,
\end{align}
to equation \eqref{limmeandenscaled} considered at the centre of the GUE spectrum.
In the limit of large values of $y$,
\begin{align}\label{rhoUlim}
\widetilde \rho_{CUE} (y) \sim
\begin{cases}
\displaystyle{\frac{1-T}{T}\,\frac{1}{y}\, e^{-2y \frac{1-T}{T}}}, &\text{if $0<T<1$,} \\
\displaystyle{\frac{1}{y^2}} & \text{if $T=1$}.
\end{cases}
\end{align}
The rescaled radial density has an exponentially light tail if $0<T<1$, and it is heavy-tailed if $T=1$ which hints at markedly different behaviour of the extreme eigenvalues in the two cases. Reflecting on \eqref{rhoUlim}, one can convince themselves that this change occurs in an infinitesimal region near $T=1$ of width $N^{-1}$.
Such a scaling regime was earlier identified and analysed from a somewhat different angle in \cite{FI2019}. The precise relation of our analysis to one in \cite{FI2019} will be given in a separate paper \cite{FK2022b}. 

On setting 
$T=1-\frac{t}{N}$, $t>0$, one can investigate this transition region in much detail \cite{FK2022b}. For example, the smallest eigenvalue modulus of the subunitary matrices $J_{CUE}$,
\begin{align*}
x_{min}=\min_{j=1, \ldots, N} |z_j| , 
\end{align*}
converges in the limit $N\to\infty$ to a random variable $X$ whose cumulative probability distribution function is given by the series
\begin{align*}
\Pr \{ X\le x\}=  \sum_{n=1}^{\infty} (-1)^{n+1}
\frac{x^{n(n-1)} }{\prod_{k=1}^n (1-x^{2k})}\, e^{t\left(1-\frac{1}{x^{2n}}\right)}
\, , \quad 0<x<1\, .
\end{align*}
This family of probability distributions interpolates between the Fr\'echet and Gumbel distributions {{and is different from the standard family of probability distributions that characterise the extreme values in long sequence of i.i.d. random variables}}.
 In the limit of small values of $t$
\begin{align*}
\lim_{t\to 0+} \Pr \left\{ X < y\sqrt{t}  \right\} = e^{- y^{-2}}\, , \quad y>0,
\end{align*}
whereas
\begin{align*}
\lim_{t\to +\infty} \Pr \left\{ 2t(1-X) -\log t +\log(\log t) < y  \right\} = e^{- e^{-y}}\, , \quad y>0.
\end{align*}


\section{Mean density of eigenvalues at finite matrix dimensions}
\label{Sec:4}

Our analysis of various scaling regimes of the random matrix ensemble \eqref{defens} -- \eqref{GUE} is based on finite-$N$ expressions for the mean eigenvalue density and the mean density of imaginary parts in terms of orthogonal polynomials, see equations \eqref{density_N} -- \eqref{density_exact} and \eqref{denim} -- \eqref{F}. These representations are new and the current Section contains their derivations.

\subsection{Joint eigenvalue density and correlation functions}
Our starting point is a closed form expression for the joint density $P_N(z_1, \ldots, z_n)$ of the eigenvalues $z_k=X_k+iY_k$ of  $J$ \eqref{defens} -- \eqref{GUE}:
\begin{align}
\label{JPD}
	P_N(z_1, \ldots, z_N) & = \\[1ex]
	\nonumber
	\MoveEqLeft[6]
	\frac{N^{N^2/2}}{\left(2\pi\right)^{N/2}
		N! G(N) \gamma^{N-1}}  \,
	\exp{-\frac{N}{2} \left(\gamma^2 + \sum_{k=1}^N \re \left(z_k^2\right)\right)}
	\,\,  \delta \left(\gamma- \sum_{k=1}^N \im z_k\right) \prod\limits_{j<k}^{N}\left|z_j-z_k\right|^2\, ,
\end{align}
 where $G(N)$ is the Barnes $G$-function.
 This expression was derived in \cite{FyodKhor_systematic_1999} (see also \cite{Kozhan17}) and, for the obvious reason, it holds for $(z_1, \ldots, z_N)\in \myC_{+}^N$, where $\myC_+$ is the upper half of the complex plane $\myC_+=\{z=X+iY: \, Y\ge 0 \}$

The first key fact that makes our analysis possible is that the eigenvalue correlation functions
 \begin{align*}
R_{N,n}\left(z_1, \ldots,z_n\right) =
\dfrac{N!}{\left(N-n\right)!} \intd\limits_{\myC_+^{N-n}}
P_N (z_1,\ldots,z_n,z_{n+1},\ldots,z_{N}) \prodd\limits_{k=n+1}^N d X_k dY_k,
\end{align*}
can be expressed in terms of averages of products of characteristic polynomials of random matrices $J(\widetilde \gamma)$ having the same structure as \eqref{defens} -- \eqref{GUE} but of smaller dimension and with a different parameter $\gamma$. The relevance of this to our investigation is in that the mean eigenvalue density $\rho_N(X,Y)$ \eqref{meandenjoint} which is the main object of our interest is
 \begin{align}\label{rhoR}
\rho_N(X,Y) = \frac{1}{N}R_{N,1}(X+iY)\, .
\end{align}
It has been shown in \cite{FyodKhor_systematic_1999} that
\begin{align*}
	R_{N,n}\left(z_1, \ldots,z_n\right) = &
	 \dfrac{1}{\left(2\pi\right)^{n/2}\gamma^n} \left(1-\frac{\sum_{k=1}^n Y_k}{\gamma}\right)^{N-n-1}
	\dfrac{N^{\frac{n^2}{2}}(N-n)^{Nn-n^2}}{\prod\limits_{j=1}^{n}
		\left(N-j-1\right)!}\,
		\prod_{1\le j<k\le n}\left|z_j-z_k\right|^2 \times
		\\[1ex]
		    \MoveEqLeft[3]
	\exp{-\frac{N}{2}\sum_{k=1}^{n}X_k^2-N\sum_{k=1}^n Y_k\left(\gamma -Y_k\right)}
	\left\langle \,\,
	\prod_{k=1}^n
	\left|\det
	\left[\widehat{z}_k\mathbf{1}_{N-n}-
	J_
	{\widehat{\gamma}-\sum_{k=1}^n\widehat{Y}_k}
	\right]
	\right|^2\right\rangle_{\!\!H_{N-n}},
\end{align*}
where
\begin{align*}
\widehat{\gamma} =  \left(\frac{N}{N-n}\right)^{\!1/2} \gamma, \quad \quad
\widehat{z}_k  = \left(\frac{N}{N-n}\right)^{\!1/2} (X_k+iY_k), \quad \widehat{Y}_k  = \left(\frac{N}{N-n}\right)^{\!1/2} Y_k\, ,
\end{align*}
and $J_{\widehat{\gamma}-\sum_{k=1}^n\widehat{Y}_k } $ are the random matrices \eqref{defens} -- \eqref{GUE} of dimension $N-n$ with $N$ in \eqref{GUE} replaced by $N-n$ and $\gamma$ in \eqref{defens_Gamma} replaced by 
$\widehat{\gamma}-\sum_{k=1}^n\widehat{Y}_k$,
\begin{align*}
J_{\widehat{\gamma}-\sum_{k=1}^n\widehat{Y}_k } =
H_{N-n} + i \left( \widehat{\gamma}-\sum_{k=1}^n\widehat{Y}_k \right) \diag (1, 0, \ldots, 0)\, .
\end{align*}

\smallskip

The GUE average $\left\langle \ldots \right\rangle_{H_{N-n}} $ of the product of the characteristic polynomials of $J_{\widehat{\gamma}-\sum_{k=1}^n\widehat{Y}_k}$ can be performed with the help of the following proposition which we prove in Appendix.

\begin{proposition}\label{p:chpol} Let
\begin{align*}
F_{\gamma}\left(z_1,z_2,\ldots,z_n\right) =
\left\langle \prodd\limits_{j=1}^n
\left|\det\left(z_j\mathbf{1}_N-J_{\gamma}\right)
\right|^2\right\rangle\, ,
\end{align*}
where $J_{\gamma}$ are the rank-one deviations from the GUE of dimension $N$ defined by \eqref{defens} -- \eqref{GUE} and the average is taken over the GUE distribution \eqref{GUE}. Then
	\begin{align*}
	F_{\gamma}(z_1,z_2,\ldots,z_n) &=
	\\[1ex]  \MoveEqLeft[5]
	\frac{1}{2^n} 
	\, \left(\frac{N}{\pi}\right)^{\!\!2n^2}
	\!\!\! \intd D[S_{2n}] \, \exp{-\frac{N}{2}\Tr S_{2n}^2} \,
	{\det}^{N-1}\left(Z_{2n}+iS_{2n}\right)
	\,
	\det \left(Z_{2n}+iS_{2n}
	-i\gamma L_{2n}\right),
	\end{align*}
	where the integration is over the space of  $2n\times 2n$ Hermitian matrices $S_{2n}$,  $D[S_{2n}]$ is the
	standard volume element in this space and
	\begin{align*}
	Z_{2n}=\diag \left(z_1,z_2,\ldots,z_n,\overline{z}_1,
	\overline{z}_2,\ldots,\overline{z}_n\right), \quad\quad  L_{2n}=\diag
	\left(1,-1\right)\otimes \mathbf{1}_n.
	\end{align*}
\end{proposition}

Using this Proposition one arrives, after rescaling $S_{2n} = \left(\frac{N}{N-n}\right)^{\!1/2} \widehat{S}_{2n}$ in the resulting matrix integral, at a useful integral representation for the eigenvalue correlation functions in the random matrix ensemble  \eqref{defens} -- \eqref{GUE}:
\begin{align}  \label{corrint}
	R_{N,n}\left(z_1, \ldots,z_n\right) =
	& \\[1ex]
	\MoveEqLeft[6]
	\nonumber
	 \frac{c_N}{\gamma^n} \left(1-\frac{\sum_{k=1}^n Y_k}{\gamma}\right)^{N-n-1}
		\exp{-\frac{N}{2}\sum\limits_{k=1}^n X_k^2 - N\sum\limits_{k=1}^n Y_k\left(\gamma-Y_k\right)}
		\prodd\limits_{1\le j<k\le n}\left|z_j-z_k\right|^2  \times
	\\[1ex] \nonumber
	 \MoveEqLeft[6]
	 \intd d[\widehat{S}_{2n}]  \exp{-\frac{N}{2}\Tr \widehat{S}_{2n}^{\,2}}
	{\det}^{N-n-1}\left[Z_{2n}+i\widehat{S}_{2n}\right]
	\det \left[Z_{2n}+i\widehat{S}_{2n}
	-i\left(\gamma-\sum_{k=1}^n Y_k \!\right) L_{2n}\right]
	\end{align}
with
\begin{align*}
c_{N,n}(\gamma) = 	 \dfrac{N^{3n^2/2+Nn}}{(2\gamma)^n\, (2\pi)^{n/2}\, \pi^{2n^2}
	\prod\limits_{j=1}^{n}
		\left(N-j-1\right)!
	}
\, .
\end{align*}

\subsection{Mean Density of complex eigenvalues}
Setting $n=1$ and $z_1=X+iY$  in \eqref{corrint} and then shifting the variable of integration by making the substitution $\widehat{S}_2 = S_2-Y L_2$ in the matrix integral, one obtains the following integral representation for the mean density of eigenvalues \eqref{rhoR} in the random matrix ensemble \eqref{defens} -- \eqref{GUE}:
	\begin{align} \label{densint}
		&\rho_N(X,Y)
		=
		\dfrac{1}{2\gamma\pi^{2}\sqrt{2\pi}} \dfrac{N^{N+1/2}}{\left(N-2\right)!}\,
		 \left(1-\frac{Y}{\gamma}\right)^{N-2}
		\exp{-\frac{N}{2}X^2-N Y\gamma} \times
		\\[1ex] \nonumber
		&		
		\intd D[S_2]\, \exp{-\frac{N}{2}\Tr S_2^2+N Y\Tr S_2L_2}
		{\det}^{N-2} (X \mathbf{1}_2+iS_2)
		\det \big(X \mathbf{1}_2+iS_2-i(\gamma-Y)L_2\big).
	\end{align}
It is convenient to parametrise the hermitian matrix $S_2$ by diagonalising it:
	\begin{align*}
		S_2 = U_2\Sigma_2 U_2^*,\quad \Sigma_2 = \diag \left(\sigma_1,\sigma_2\right), \quad
		\sigma_1\geq\sigma_2 \in \R,
	\end{align*}
where $U_2$ is a  $2\times 2$ unitary matrix, which can
be parametrised as
	\begin{align*}
		U_2 = \left(
			\begin{array}{cc}
				\cos\theta & \sin\theta e^{i\phi} \\
				-\sin\theta e^{-i\phi} & \cos\theta
			\end{array}
		\right), \quad \theta \in \left[0,\frac{\pi}{2}\right],\,
		\phi \in \left[0,2\pi\right].
	\end{align*}
Noting that
	\begin{align*}
	D[S_2] =\left(\sigma_1-\sigma_2\right)^2\,  \frac{\sin (2\theta) }{2}\, d\sigma_1\, d\sigma_2\,  d\theta \, d\phi.
	\end{align*}
one arrives, on making the substitution $S_2 = U_2\Sigma_2 U_2^*$ in \eqref{densint}, at
	\begin{align*}
		\rho_N(X,Y) = &  \frac{c_N}{\gamma}
		\left(1-\frac{Y}{\gamma}\right)^{N-2}
		\exp{-\frac{N}{2}\left(X^2+2Y\gamma\right)}  \times
         \\[1ex]
          \MoveEqLeft[5]
		\int_{0}^{\frac{\pi}{2}} \! d\theta\!
		\int_{-\infty}^{+\infty} \!\!d\sigma_1 \!\int_{-\infty}^{\sigma_1}\!\! d\sigma_2\,
		(\sigma_1-\sigma_2)^2 \sin (2\theta )
		\,
		\exp{-\frac{N}{2}\left(\sigma_1^2+\sigma_2^2\right)
			+N Y\left(\sigma_1-\sigma_2\right)\cos (2\theta) } \,
		\times
	\\[1ex]
		 \MoveEqLeft[5]
		 \left(X+i\sigma_1\right)^{N-2}
		\left(X+i\sigma_2\right)^{N-2}  \left[
		\left(X+i\sigma_1\right)
		\left(X+i\sigma_2\right)
		+\left(\gamma-Y\right)^2 -
		\left(\gamma-Y\right)\left(\sigma_1-\sigma_2\right)\cos (2\theta)
		\right],
	\end{align*}
where we have introduced
\begin{align}\label{cN}
c_N = 	\dfrac{1}{\left(2\pi\right)^{3/2}}
\dfrac{N^{N+1/2}}{\left(N-2\right)!}\sim
\dfrac{N^2e^N}{2\pi^2} \quad \quad (N\to \infty).
\end{align}
The integral over $\theta$ can be performed by the substitution $t=(
\sigma_1-\sigma_2)\cos(2\theta)$. 
This yields
	\begin{align}\label{density_N}
		\rho_N(X,Y)
=& \dfrac{c_N }{2N Y\gamma} \left(1-\frac{Y}{\gamma}\right)^{N-2} \!\!\!
		\exp{-\frac{N}{2}\left(X^2+2\gamma Y-2Y^2\right)}	J_{N}(X,Y)\, ,
\end{align}
where
	\begin{align*}
			J_{N}(X,Y)=&\int_{-\infty}^{+\infty} \!\!\!d\sigma_1\int_{-\infty}^{+\infty} \!\!\! d\sigma_2	
			\, 	
		e^{-\frac{N}{2}(\sigma_1^2+\sigma_2^2)}
		(z+ i\sigma_1)^{N-2}
		(\overline{z}+i\sigma_2)^{N-2} 	
		\dfrac{(z+i\sigma_1)
		-(\overline{z}+i\sigma_2)}{i}
		 \times 		
		\\[1ex]
		&
		\left[		
		(z+i\sigma_1)
		(\overline{z}+i\sigma_2)
		+(\gamma-Y)^2
		+\frac{\gamma-Y}{Ny}
		-(\gamma-Y)
		\dfrac{(z+i\sigma_1)
			-(\overline{z}+i\sigma_2)}{i}
		\right]\,,
	\end{align*}
with $z=X+iY$.

Further, introducing functions
\begin{align}\label{Hermint}
\pi_m(z)=\int_{-\infty}^{+\infty} \!\!d\sigma \, e^{-\frac{N}{2}\sigma^2}\, (z+i\sigma)^{N-m}\,,  \quad m=0, 1, \ldots, N,
\end{align}
one can rewrite the integral $J_{N}(X,Y)$ in the following form
\begin{align}\label{intden1}
		J_{N}(X,Y)=&-i\left[\pi_0(Z)\pi_1(\overline{z})-\pi_0(\overline{z})\pi_1(z)\right]
		\\ \nonumber
		&
 -i\left((\gamma-Y)^2
		+\frac{\gamma-Y}{Ny}\right)
		\left[\pi_1(z)\pi_2(\overline{z})-\pi_1(\overline{z})\pi_2(z)\right]&
\\ \nonumber
&+(\gamma-Y)\left[\pi_0(z)\pi_2(\overline{z})+\pi_0(\overline{z})\pi_2(z)-2\pi_1(z)\pi_1(\overline{z})\right]\, ,
\end{align}
Now one observes that $\pi_m(z)$ are actually a rescaled version of Hermite polynomials. We have that
	\begin{align}\label{Hermitpoly}
	\pi_m(z)= \sqrt{\pi}\left(\dfrac{2}{N}\right)^{\!\!\frac{N-m+1}{2}}
	\!\!
	\widetilde H_{N-m}\left(z\sqrt{\frac{N}{2}}\right)
	= \sqrt{2\pi}\pi^{1/4} \sqrt{\dfrac{\left(N-m\right)!}{N^{N-m+1}}} \, p_{N-m}\left( z\sqrt{\frac{N}{2}}\right)\, ,
	\end{align}
where $\widetilde H_k(z)$ are the monic Hermite polynomials
	\begin{align*}
	\widetilde H_k(z) =
		\left(-\frac{1}{2}\right)^{\!k} e^{\,z^2} \frac{d}{dz}\,  e^{-z^2}
	\end{align*}
and $p_k(z)$ are the orthonormal Hermite polynomials
	\begin{align*}
	p_k (z)  =
		\sqrt{\dfrac{2^k}{k!\sqrt{\pi}}} \,\,
		\widetilde H_k\left(z\right)
	\end{align*}
satisfying the orthogonality relations
	\begin{align*}
		\int_{-\infty}^{+\infty} \!\!dz \,  p_k (z)p_m (z) \, e^{-z^2} dz  =\delta_{k,m}.
	\end{align*}
The  polynomials $p_k (z)  $ also satisfy the recurrence relation
	\begin{align*}
	p_{k+1}\left(z\right) = z\sqrt{\frac{2}{k+1}}p_k\left(z\right)-
	\sqrt{\frac{k}{k+1}}p_{k-1}\left(z\right).
	\end{align*}
Using the above definitions and the expression for the eigenvalue density $\rho_N(X,Y)$ in \eqref{density_N} and with the notation $z=X+iY$ we obtain
\begin{align*}
	\rho_N(X,Y)=& \frac{N-1}{\sqrt{2N}Y\gamma} \left(1-\frac{Y}{\gamma}\right)^{N-2}
	\exp{-\frac{N}{2}X^2-N Y\left(\gamma-Y\right)} \times
	\\[1ex]
	&
	\left\{\im p_{N}\left(z\sqrt{\frac{N}{2}}\right)
	p_{N-1}\left(\overline{z}\sqrt{\frac{N}{2}}\right)
	- \left(\gamma-Y\right)\left|p_{N-1}\left(z\sqrt{\frac{N}{2}}\right)\right|^2
	\right.
	\\[1ex]
	&
	\left.+\sqrt{\frac{N}{N-1}}\left[
	\left( \left(\gamma-Y\right)^2+\dfrac{\gamma-Y}{N Y}\right)
	\im p_{N-1}\left(z\sqrt{\frac{N}{2}}\right)
	p_{N-2}\left(\overline{z}\sqrt{\frac{N}{2}}\right) \right.\right.
	\\[1ex] & \left.\left.
	+\left(\gamma-Y\right)	\re p_{N}\left(z\sqrt{\frac{N}{2}}\right)
	p_{N-2}\left(\overline{z}\sqrt{\frac{N}{2}}\right)
	\right] \right\}.
\end{align*}	
which, by using the recurrence relation, can be further rewritten as
\begin{align} \label{density_exact}
	\rho_N(X,Y)
	=&
	\frac{1}{Y\gamma}\sqrt{\frac{N}{2}} \left(1-\frac{Y}{\gamma}\right)^{N-2}
	\exp{-\frac{N}{2}X^2-N Y\left(\gamma-Y\right)} \times
	\\[1ex] \nonumber
	&
	\left\{
	\im p_{N}\left(z\sqrt{\frac{N}{2}}\right)
	p_{N-1}\left(\overline{z}\sqrt{\frac{N}{2}}\right)
	\left[
	1-\frac{1}{N}+\left(\gamma-Y\right)\left(\gamma+\frac{1}{N Y}\right)
	\right]
	\right.
	\\[1ex] \nonumber
	&
	\left.
	-
	\left| p_{N-1}\left(z\sqrt{\frac{N}{2}}\right)\right|^2
	\left[
	Y\left(\gamma-Y\right)^2+\left(\gamma-Y\right)-Y\left(\gamma-Y\right)
	\right]
		\right.
	\\[1ex]
	&
	\left.
	-
	\left|p_{N}\left(z\sqrt{\frac{N}{2}}\right)\right|^2
	\left(\gamma-Y\right)
	+	
	\re p_{N}\left(z\sqrt{\frac{N}{2}}\right)
	p_{N-1}\left(\overline{z}\sqrt{\frac{N}{2}}\right)
	X\left(\gamma-Y\right)
	\right\}.\nonumber
\end{align}

\subsection{Density of the imaginary parts}
In this section we present the derivation of the density for the imaginary parts of the eigenvalues,
irrespective of their real parts, as defined in (\ref{meandenim}).
We start with an observation, see integral 7.377 in \cite{GR2007}:
\begin{lemma}
	Let $\beta \geq \alpha$ be two non-negative integers
	and $z=X+iY$.
	Then
	\begin{align*}
		\intd\limits_{-\infty}^{\infty} e^{-\frac{N}{2}X^2}
		p_{N-\alpha}\left(z\sqrt{\frac{N}{2}}\right)
		p_{N-\beta}\left(\overline{z}\sqrt{\frac{N}{2}}\right) dX
		= \\
		\MoveEqLeft[+5]
		 i^{\beta-\alpha}\sqrt{\frac{2}{N} \frac{N^{\beta-\alpha}}{\left(N-\alpha\right)\ldots
		\left(N-\beta+1\right)}}Y^{\beta-\alpha}L^{\left(\beta-\alpha\right)}_{N-\beta}\left(-NY^2\right),
	\end{align*}
	where $L^{\left(\alpha\right)}_M$ is a standard Laguerre polynomial.
\end{lemma}
\noindent
Integrating with respect to $X$ expression for the density
	$\rho_N(X, Y)$ in (\ref{density_exact}) one gets the probability density of imaginary parts in the form
	\begin{align}\label{denim}
		\rho^{(\im)}_N\left(Y\right) = \dfrac{1}{Y\gamma}\left(1-\frac{Y}{\gamma}\right)^{N-2}e^{-NY\left(\gamma-Y\right)}F_N(Y)
\end{align}
with
\begin{align}
\label{F1}	F_N(Y)=&
		\frac{N-1}{N}YL^{\left(1\right)}_{N-1}\left(-NY^2\right)
		-\frac{N-1}{N}\left(\gamma-Y\right)L^{\left(0\right)}_{N-1}\left(-NY^2\right)&
		\\ \nonumber
		&		
		+Y\left[\left(\gamma-Y\right)^2+\frac{\gamma-Y}{NY}\right]L^{\left(1\right)}_{N-2}\left(-NY^2\right)
		-\left(\gamma-Y\right)Y^2
		L^{\left(2\right)}_{N-2}\left(-NY^2\right)&		
		\\
\label{F2} =&-2\gamma L^{\left(0\right)}_{N-1}\left(-NY^2\right)
+\left(\frac{N-1}{N}3Y+\frac{2\gamma}{N}\right)L^{\left(1\right)}_{N-1}\left(-NY^2\right)&\\ \nonumber
&+\left[-2Y+Y(\gamma-Y)^2\right]L^{\left(1\right)}_{N-2}\left(-NY^2\right)&
\\
\label{F}
		=&		
			\frac{N-1}{N}\left(3Y-2\gamma\right)L^{\left(1\right)}_{N-1}\left(-NY^2\right)
			+\left[2\gamma-2Y+Y\left(\gamma-Y\right)^2\right]
			L^{\left(1\right)}_{N-2}\left(-NY^2\right)&
		\end{align}
where we systematically used the recursion relations:
\begin{align*}
L_{N-1}^{\left(0\right)}\left(-NY^2\right)=L_{N-1}^{\left(1\right)}\left(-NY^2\right)-L_{N-2}^{\left(1\right)}\left(-NY^2\right)
\end{align*}
and
\begin{align*}
-Y^2L_{N-2}^{\left(2\right)}\left(-NY^2\right)=L_{N-2}^{\left(1\right)}\left(-NY^2\right)-\frac{N-1}{N}
L_{N-1}^{\left(1\right)}\left(-NY^2\right).
\end{align*}

\section{Proof of Theorems \ref{Thm_App 1} and \ref{Thm_App 1a}}
\label{Proofs}

In both proofs we use the following integral representation for the Laguerre polynomials in terms of the modified
Bessel functions $I_{\alpha}(x)$ (see, e.g. equation 4.19.13 in \cite{Lebedevbook}):
\begin{align}\label{Lagint}
		L_{N-k}^{\left(\alpha\right)}\left(-NY^2\right) =
		\frac{2 N^{N-k+1}}{\left(N-k\right)!}\frac{e^{-NY^2}}{|Y|^{\alpha}}
	\intd\limits_{0}^{\infty}  \tau^{2N-2k+\alpha+1}e^{-N\tau^2} I_{\alpha}(2\tau |Y|N)\, d \tau \quad (\alpha >-1).
\end{align}
The integral in \eqref{Lagint} can be evaluated in the limit $N\to\infty $ in various scaling regimes for $Y$ using the Laplace method, see Appendix B. The resulting asymptotic expression depends on the scaling of the variable $Y>0$ with $N$.

\begin{proof}[\bf Proof of Theorem \ref{Thm_App 1}]
Consider the scaling regime \eqref{EValues} with $\gamma >0$ being fixed. In this regime the asymptotic form of the mean density of the imaginary parts can be found using the leading order form of $L_{N-k}^{\left(1\right)}\left(-NY^2\right) $ which can be  read from (\ref{AsyLagg}) as
	\begin{align}\label{AsyLagg0}
		L_{N-k}^{\left(1\right)}\left(-NY^2\right) \sim
		\frac{e^{NYr_*}}{\sqrt{2\pi N}}\frac{r_*(Y)^{-2(N-k+1)}}
		{Y^{3/2}\left(Y^2+4\right)^{1/4}}, \quad r_* (Y)= \frac{\sqrt{Y^2+4}-Y}{2}\, .
	\end{align}

On substituting (\ref{AsyLagg0}) into  (\ref{F}) one gets an asymptotic expression for the density (\ref{denim}) precisely in the Large Deviation form (\ref{LDform}) with the rate function (\ref{LDrate}) and the pre-exponential factor in the form
\begin{align}\label{LDpreexp}
 \frac{1}{\sqrt{N}}\Psi_{\gamma}(Y)&=\frac{1}{\sqrt{2\pi N}}\frac{\gamma}{(\gamma-Y)^2}\frac{3Y-2\gamma+r_*(Y)^2(\gamma-Y)\left(2+Y(\gamma-Y)\right)}{Y^{5/2}(Y^2+4)^{1/4}}\, .
\end{align}
Finally, by exploiting the relation $1-r_*(Y)^2=Yr_*(Y)$,
\begin{align*}
3Y-2\gamma+r_*(Y)^2(\gamma-Y)\left(2+Y(\gamma-Y)\right)=& Y-(\gamma-Y)\left[2(1-r_*(Y)^2)-r_*(Y)^2\,Y(\gamma-Y)\right]  \\
=&Y-(\gamma-Y)\left[2r_*(Y)Y-r_*(Y)^2\,Y(\gamma-Y)\right] \\
=& Y\left[1-r_*(Y)(\gamma-Y)\left(2-r_*(Y)(\gamma-Y)\right)\right].
\end{align*}
This brings the  function $\Psi_{\gamma}(Y)$ in \eqref{LDpreexp} to the form as given in (\ref{LDpreexp1}).

\medskip

To analyse the shape of the rate function $\Phi_{\gamma}(Y)$ in (\ref{LDrate}) it is convenient to parametrise
\begin{align}\label{param1}
Y=e^{\theta}-e^{-\theta}, \quad \theta>0\, .
\end{align}
In this parametrisation, the rate function transforms to
\begin{align*}
\widetilde \Phi_{\gamma}(\theta):= \Phi_{\gamma}\big(e^{\theta}-e^{-\theta}\big)=\gamma\left(e^{\theta}-e^{-\theta}\right)+1-e^{2\theta}-2\theta-\ln{\left(1-\frac{e^{\theta}-e^{-\theta}}{\gamma}\right)}\, ,
\end{align*}
and its derivative in $\theta$ factorises as follows:
\begin{align*}
\widetilde \Phi^{\prime}_{\gamma}(\theta)=& \gamma\left(e^{\theta}+e^{-\theta}\right)-2\left(e^{2\theta}+1\right)
+\frac{e^{\theta}+e^{-\theta}}{\gamma-\left(e^{\theta}-e^{-\theta}\right)}\\
=&\left(e^{\theta}+e^{-\theta}\right)\left(\gamma-e^{\theta}\right)
\left[1-\frac{e^{\theta}}{\gamma-\left(e^{\theta}-e^{-\theta}\right)}\right]\, .
\end{align*}
Therefore, the stationary points of $\widetilde \Phi_{\gamma}(\theta)$ solve the equations
\begin{align}\label{eq1b}
e^{\theta}=\gamma
\end{align}
and
\begin{align}\label{condition}
e^{\theta}=\gamma-\left(e^{\theta}-e^{-\theta}\right)\, .
\end{align}
These equations yields two stationary points $e^{\theta_*}=\gamma$ and $e^{\theta_{**}}= \frac{\gamma+\sqrt{8+\gamma^2}}{4}$. Correspondingly, the rate function $\Phi_{\gamma}(Y)$ has two stationary points
\begin{align*}
Y_*=\gamma-\gamma^{-1} \quad \text{ and } \quad Y_{**}=\frac{3\gamma-\sqrt{8+\gamma^2}}{4}=\frac{2\left(\gamma-\frac{1}{\gamma}\right)}{3+\sqrt{1+8/\gamma^2}}\, .
\end{align*}

It is evident that if $0<\gamma<1$ both stationary points $Y_*$ and $Y_{**}$ are negative. One can easily check that in this case 
$\Phi_{\gamma}(Y) $ is monotonically increasing on the interval $Y>0$ and is positive on this interval.

If $\gamma>1$ then
taking the second derivative in $\theta$ one can easily show that
\begin{align*}
\widetilde \Phi^{\prime\prime}_{\gamma}(\theta_*)=(\gamma^2-1)(\gamma^2+1)>0, \quad
\widetilde \Phi^{\prime\prime}_{\gamma}(\theta_{**})=-\frac{\left(e^{2\theta_{**}}-e^{-2\theta_{**}}\right)
 \left(1+\gamma e^{\theta_{**}}\right)}{\left[\gamma-\left(e^{\theta_{**}}-e^{-\theta_{**}}\right)\right]^2} <0,
\end{align*}
so that $Y_*$ is the point of local minimum of the rate function $\Phi_{\gamma}(Y)$, and  $Y_{**}$ is the point of local maximum. It is also easy to verify that the rate function $\Phi_{\gamma}(Y)$ vanishes in the limit $Y\to 0$ and also at $Y=Y_{*}$, staying positive at all other $Y> 0$, so that that the point $Y=Y_*$ is the point of absolute minimum. Finally, to verify that the pre-exponential factor (\ref{LDpreexp1}) vanishes at
 $Y=Y_{**}$
  it suffices to show that  $r_*(Y_{**})(\gamma-Y_{**})=1$. On noticing that
 \begin{align*}
r_*(Y)=\frac{\sqrt{Y^2+4}-Y}{2}=e^{-\theta}\, .
\end{align*}
this relation evidently follows from  (\ref{param1}) and (\ref{condition}).
\end{proof}

\begin{proof}[{\bf Proof of Theorem \ref{Thm_App 1a}}]
In the scaling regime \eqref{sr} the variable $Y$ scales with $N^{-1/3}$. As $NY \gg 1$ in this case, the required asymptotic expressions for Laguerre polynomials can be read from (\ref{AsyLagg}).
It turns out that in order to calculate the density of imaginary parts to leading order in this regime, one has to retain the subleading term in the pre-exponential factor as specified in (\ref{AsyLagg}). On substituting $Y=m N^{-1/3}$  in (\ref{AsyLagg}) we obtain that with the required precision
\begin{align}\label{AsyLagga}
		L_{N-1}^{\left(1\right)}\left(-N^{1/3}m^2\right) &=
		\frac{e^{N\mathcal{L}_0(Y)}}{\sqrt{2\pi m^3}\, \left(Y^2+4\right)^{1/4}} \left(1-\frac{3}{16}\frac{1}{mN^{2/3}}\right)\\ \label{AsyLaggb}
L_{N-2}^{\left(1\right)}\left(-N^{1/3}m^2\right) &=
		\frac{e^{N\mathcal{L}_0(Y)}}{\sqrt{2\pi m^3}\, \left(Y^2+4\right)^{1/4}} \, r_*(Y)^2\left(1-\frac{3}{16}\frac{1}{mN^{2/3}}\right)\\
\label{AsyLaggc}
L_{N-1}^{\left(0\right)}\left(-N^{1/3}m^2\right) &=
		\frac{e^{N\mathcal{L}_0(Y)}}{N^{1/3}\sqrt{2\pi m^3}\, \left(Y^2+4\right)^{1/4}} \left(1+\frac{1}{16}\frac{1}{mN^{2/3}}\right)\, ,
\end{align}
where $\mathcal{L}_0(Y)=Yr_*(Y)-2\ln{r_*}(Y)$ with $r_*$ \eqref{rstar} and $\ln r_*$ expanded in powers of $Y\ll 1$:
\begin{align}\label{expanr}
r_*(Y)=1-\frac{Y}{2}+\frac{Y^2}{8}+O(Y^4), \quad \ln{r_*(Y)}=-\frac{Y}{2}+\frac{Y^3}{48}+O(Y^4).
\end{align}
It is easy to see that the overall exponential behaviour of the mean density \eqref{denim} will still  be given by  (\ref{LDrate}) duly expanded:
\begin{align}\label{small_exp}
\Phi_{\gamma}(Y)=Y\left(\gamma+\frac{1}{\gamma}-2\right)-\frac{Y^2}{2}\left(1-\frac{1}{\gamma^2}\right)
+\frac{Y^3}{3}\left(\frac{1}{\gamma^3}-\frac{1}{4}\right)+O(Y^4)\, .
\end{align}
  Putting in here the scaling form $\gamma=1+\frac{\alpha}{N^{1/3}}$  and recalling $Y=\frac{m}{N^{1/3}}$ we find from (\ref{small_exp}), assuming that the parameters $\alpha\in \mathbb{R}$ and $m>0$ are fixed, that
\begin{align*}
N\Phi_{1+\frac{\alpha}{N^{1/3}}}\left(\frac{m}{N^{1/3}}\right)=m\alpha^2-m^2\alpha+\frac{m^3}{4}=
m\left(\alpha-\frac{m}{2}\right)^2:=\Phi_{\alpha}(m)\,.
\end{align*}
This verifies the exponent in  (\ref{scaleform}).
To find the pre-exponential terms we find it most convenient to use equation (\ref{F2}).
Substituting there
(\ref{AsyLagga})-(\ref{AsyLaggc}) we first get
\begin{align}\label{eq:mf}
F_N(m)=& \frac{e^{N\mathcal{L}_0(Y)}}
{\sqrt{2\pi m}\, \left(Y^2+4\right)^{1/4}}\left\{
-\frac{2\gamma}{N^{1/3}}r_*(Y) \left(1+\frac{1}{16}\frac{1}{mN^{2/3}}\right)
\right. \\ \nonumber & \left.
 +\frac{1}{m}
\left(1-\frac{3}{16}\frac{1}{mN^{2/3}}\right)\left(\frac{3m}{N^{1/3}}+\frac{2\gamma}{N}-\frac{3m}{N^{4/3}}\right)\right.
\\ \nonumber
& \left.+\left[-\frac{2}{N^{1/3}}+\frac{1}{N^{1/3}}\left(\gamma^2-2\gamma\frac{m}{N^{1/3}}+\frac{m^2}{N^{2/3}} \right)\right] r_*(Y)^2\left(1-\frac{3}{16}\frac{1}{mN^{2/3}}\right)\right\}.
\end{align}
After rearranging and collecting the relevant terms in the above expression we arrive at
 \begin{align}\label{2terms}
F_N(m) = & \frac{e^{N\mathcal{L}_0(Y)}}
{\sqrt{2\pi m}\, \left(Y^2+4\right)^{1/4}}\left\{\frac{(r_*(Y)\gamma-1)^2+2(1-r_*(Y)^2)}{N^{1/3}}-\frac{2\gamma m r_*(Y)^2}{N^{2/3}} \right.
\\
\nonumber
\MoveEqLeft[-5]
\left.+\frac{1}{N}\left[-\frac{9}{16m}+m^2r_*(Y)^2-\frac{3}{16}\frac{(\gamma^2-2)r_*(Y)^2}{m}+\frac{2\gamma}{m}-\frac{\gamma
r_*(Y)}{8m}\right]\right\}.
\end{align}
The expansion (\ref{expanr}) together with $\gamma=1+\frac{\alpha}{N^{1/3}}$ give the relations
\begin{align}\label{eq:aa}
\frac{(r_*(Y)\gamma-1)^2+2(1-r_*(Y)^2)}{N^{1/3}}=&\frac{2m}{N^{2/3}}+
\frac{1}{N}\left[\left(\alpha-\frac{m}{2}\right)^2-m^2\right]
 \end{align}
and
 \begin{align}\label{eq:ab}
 -\frac{2\gamma m r_*(Y)^2}{N^{2/3}}=& -\frac{2m}{N^{2/3}}-\frac{2m(\alpha-m)}{N}\, ,
 \end{align}
which are exact to the subleading order. We can now see that the leading order terms inside the curly brackets in (\ref{2terms}) cancel. 
This also implies that at the leading order it is enough to replace the factor $\left(Y^2+4\right)^{1/4}$ in (\ref{2terms}) with $\sqrt{2}$. Finally, adding the leading order contribution from
 \begin{align*}
\frac{1}{N}\left[-\frac{9}{16m}+m^2r_*(Y)^2-\frac{3}{16}\frac{(\gamma^2-2)r_*(Y)^2}{m}+\frac{2\gamma}{m}-\frac{\gamma
r_*(Y)}{8m}\right]=\frac{1}{N}\left(m^2+\frac{3}{2m}\right)
\end{align*}
to the $1/N$ terms in \eqref{eq:aa} --\eqref{eq:ab} results in
\begin{align}\label{Ffin}
F_N(m)=
\frac{e^{N\mathcal{L}_0(Y)}}
 {2\sqrt{\pi m}}\frac{1}{N}\left[\frac{3}{2m}+\left(\alpha-3\frac{m}{2}\right)^2\right]\, ,
\end{align}
thus verifying the pre-exponential factors in  (\ref{scaleform}).
\end{proof}

\medskip

Let us finally present the derivation of the marginal density of imaginary parts (\ref{denimy1}) pertinent
to keeping  the product $y=YN$ fixed as $N\to \infty$. This task is straightforwardly achieved by performing the limit $N\to \infty$ in (\ref{denim}) via
substituting the corresponding asymptotics of Laguerre polynomials (\ref{Lagasy2}) into  the formula (\ref{F2}) and
using the identity $\frac{d}{dy}I_1(2y)=I_0(2y)-I_2(2y)$.

\section{Proof of Theorem \ref{Thm_App 1b}}

\begin{proof} We will use equations \eqref{density_N} -- \eqref{intden1}  
which express the mean density of eigenvalues $\rho_N(X,Y)$ in terms of the rescaled Hermite polynomials $\pi_{k} (X+iY)$ \eqref{Hermitpoly}. 

Using the integral representation in \eqref{Hermint} it can be shown that in the scaling limit
 \begin{align}\label{smallz}
 z=X+iY, \quad  X=\frac{q}{N^{1/3}}, \quad Y=\frac{m}{N^{1/3}}>0
 \end{align}
 the rescaled Hermite polynomials $\pi_{k} (z)$ are given by the asymptotic equations
  \begin{align}\label{asyHer}
 \pi_k(z)\sim \sqrt{\frac{2\pi}{N(1+\sigma_{+}^2)}}\left(-i\sigma_{+}\right)^k\,
 e^{-\frac{N}{2}\left(1+iz\sigma_{+}+2\ln{\left(-i\sigma_{+}\right)}\right)}\, , \\
 \pi_k(\overline{z})\sim  \sqrt{\frac{2\pi}{N(1+\sigma_{-}^2)}}\left(-i\sigma_{-}\right)^k\,
 e^{-\frac{N}{2}\left(1+i\overline{z}\sigma_{-}+2\ln{\left(-i\sigma_{-}\right)}\right)}\, ,
 \end{align}
 where we have introduced the notations
 \begin{align}
 \sigma_{+}=\frac{iz+\sqrt{4-z^2}}{2}, \quad \sigma_{-}=\frac{i\overline{z}-\sqrt{4-\overline{z}^2}}{2}\, .
 \end{align}
 This implies for $J_{N}(X,Y)$  (\ref{intden1}) that
 \begin{align}\label{asyJ1}
 J_{N}(X,Y)\sim \frac{2\pi}{N}e^{-N}\frac{( \sigma_{+}- \sigma_{-})}{\sqrt{(1+\sigma_{+}^2)(1+\sigma_{-}^2)}}
 e^{-N\left(\frac{i}{2}\left(z\sigma_{+}+\overline{z}\sigma_{-}\right)+\ln{\left(-\sigma_{+}\sigma_{-}\right)}\right)}
 \\ \times \left[\left(1-\sigma_{+}(\gamma-y)\right)\left(1+\sigma_{-}(\gamma-Y)\right)-\sigma_{+}\sigma_{-}
 \frac{\gamma-Y}{NY}\right]\, .
 \label{asyJ}
 \end{align}
 We are here interested in the limit of small $|z|$ \eqref{smallz},
 and, hence, can use the expansions
 \begin{align*}
 \sigma_{+}=1+\frac{iz}{2}-\frac{z^2}{8}+\ldots, \quad \sigma_{-}=-1+\frac{i\overline{z}}{2}+\frac{\overline{z}^2}{8}+\ldots
 \end{align*}
 and, consequently,
 \begin{align*}
 1-\sigma_{+}(\gamma-Y) =&1-\gamma+Y+\frac{Y}{2}\gamma-\frac{iX}{2}\gamma+O(|z|^2)\, , \\
 1-\sigma_{-}(\gamma-Y)  =& 1-\gamma+Y+\frac{Y}{2}\gamma+
 \frac{iX}{2}\gamma+O(|z|^2)\, .
 \end{align*}
 Hence,
 \begin{align*}
 \left(1-\sigma_{+}(\gamma-Y)\right)\left(1+\sigma_{-}(\gamma-Y)\right)=
 \left[1-\gamma+Y\left(1+\frac{\gamma}{2}\right)\right]^2+\frac{X^2}{4}\gamma^2 +O\left( X^2+Y^2\right).
\end{align*}
Setting here $X=\frac{q}{N^{1/3}}$, $Y=\frac{m}{N^{1/3}}$ and $\gamma=1+\frac{\alpha}{N^{1/3}}$ one obtains that   to leading order in $N$
\begin{align*}
 \left(1-\sigma_{+}(\gamma-Y)\right)\left(1+\sigma_{-}(\gamma-Y)\right) =
 \frac{1}{N^{2/3}}\left[\left(\frac{3}{2}m-\alpha\right)^2+\frac{q^2}{4}\right]\, .
\end{align*}
With the same precision we have
\begin{align*}
-\sigma_{+}\sigma_{-}
 \frac{\gamma-Y}{NY} = \frac{1}{N^{2/3}}\frac{1}{m}\, ,
\end{align*}
and, consequently,
\begin{align}\label{asypref}
\left[\left(1-\sigma_{+}(\gamma-Y)\right)\!\left(1+\sigma_{-}(\gamma-Y)\right)-\sigma_{+}\sigma_{-}
 \frac{\gamma-Y}{NY}\right] =   \frac{1}{N^{2/3}}\left[\frac{1}{m}+\!\left(\frac{3m}{2}-\alpha\right)^{\!2}\!+\frac{q^2}{4}\right]\,  .
 \end{align}
On inspecting (\ref{density_N} )  and \eqref{asyJ1}, one concludes that the overall exponential factor in (\ref{density_N} ) is given by $e^{-N\widetilde{\Phi}_{\gamma}}$, where
\begin{align*}
 \widetilde{\Phi}_{\gamma}= \frac{i}{2}(z\sigma_++\overline{z}\sigma_{-})+ \ln{\left(-\sigma_{+}\sigma_{-}\right)}
 +\frac{X^2}{2}+\gamma Y -Y^2-\ln{\left(1-\frac{Y}{\gamma}\right)}.
 \end{align*}
The leading order form of $ \widetilde{\Phi}_{\gamma}$ can be found by expanding in powers of $X$ and $Y$, in a similar way as before:
 \begin{align*}
 \frac{i}{2}(z\sigma_++\overline{z}\sigma_{-})& = -Y-\frac{X^2-Y^2}{2}+\frac{Y}{8}(3X^2-Y^2)+O(|z|^4)\, ,\\
 \ln{\left(-\sigma_{+}\sigma_{-}\right)}& = -Y+\frac{Y^3}{24}-\frac{YX^2}{8}+O(|z|^4)\, ,\\
  -\ln{\left(1-\frac{Y}{\gamma}\right)}& = \frac{Y}{\gamma}+\frac{Y^2}{2\gamma^2}+\frac{Y^3}{3\gamma^3}+O(Y^4)\, .
 \end{align*}
 Adding all contributions,
\begin{align*}
 \widetilde{\Phi}_{\gamma}=Y\frac{(\gamma-1)^2}{\gamma}+\frac{Y^2}{2\gamma^2}(1-\gamma^2)+\frac{Y^3}{3}
 \left(\frac{1}{\gamma^3}-\frac{1}{4}\right)+\frac{Y X^2}{4}+O(|z|^4)\, .
\end{align*}
Setting here $X=\frac{q}{N^{1/3}}$, $Y=\frac{m}{N^{1/3}}$ and $\gamma=1+\frac{\alpha}{N^{1/3}}$, one obtains that to leading order
%
 \begin{align}\label{exponscaled}
 \widetilde{\Phi}_{\gamma}=\frac{m}{N}\left(\left(\frac{m}{2}-\alpha\right)^2+\frac{q^2}{4}\right)\, .
\end{align}
  Combining (\ref{exponscaled}) with (\ref{asypref}), and trivially taking into account asymptotic expressions for the remaining multiplicative factors in (\ref{density_N} ) and (\ref{asyJ}),
  one arrives at  (\ref{scaleformcent}).
\end{proof}

\medskip
\noindent {\bf Acknowledgments:} The authors are grateful to Guillaume Dubach for highly useful comments on the revised version of the manuscript. The authors are also grateful to Bertrand Lacroix A Chez Toine for bringing their attention to paper \cite{cond_trans_rev} and the similarity between the spectral restructure in the random matrix ensemble \eqref{defens}--\eqref{GUE} and the condensation transition in models of mass transport. Y.V.F. acknowledges financial support from EPSRC Grant EP/V002473/1 ``Random Hessians and Jacobians: theory and applications''.

\appendix

\section[\appendixname~\thesection]{Proof of Proposition~\ref{p:chpol}}
We prove here a more general version of Proposition~\ref{p:chpol} which holds for rank-$M$ deviations
 $J = H + i\Gamma, \,\,\Gamma=\diag \left\{\gamma_1,\ldots,\gamma_M,0,\ldots,0\right\}$ from the GUE \eqref{GUE}
with arbitrary real parameters $\gamma_j,  \quad j=1,\ldots,M<N$.
The Proposition~\ref{p:chpol} follows as the special case $M=1$.

\begin{proposition}\label{p:chpolM}
Let $\Gamma=\diag (\gamma_1,\ldots,\gamma_M, 0,\ldots,0)$ be a diagonal matrix of dimension $N$ with $M<N$ non-zero real entries $\gamma_j$ and
\begin{align*}
F_{\Gamma}\left(z_1,z_2,\ldots,z_n\right) =
\left\langle \prodd\limits_{j=1}^n
\left|\det\left(z_j\mathbf{1}_N -H - i\Gamma \right)
\right|^2\right\rangle_{\!\!\!H}\, ,
\end{align*}
where the average is taken over the GUE distribution \eqref{GUE}. Then
	\begin{align}\label{FGamma}
	F_{\Gamma}\left(z_1,z_2,\ldots,z_n\right) =
	\\ \nonumber
	\MoveEqLeft[+5]
	\frac{1}{2^n}\left(\frac{N}{\pi}\right)^{2n^2}
	\!\!\!
	\intd D[S_{2n}] \, e^{-\frac{N}{2}\Tr S_{2n}^2}
	{\det}^{N-M}(Z_{2n}+iS_{2n})
	\prodd\limits_{j=1}^M\det(Z_{2n}+iS_{2n}
	-i\gamma_jL_{2n}),
	\end{align}
where the integration is over the space of  $2n\times 2n$ Hermitian matrices $S_{2n}$,  $D[S_{2n}]$ is the
	standard volume element in this space and
\begin{align*}
	Z_{2n}=\diag \left(z_1,z_2,\ldots,z_n,\overline{z}_1,
	\overline{z}_2,\ldots,\overline{z}_n\right), \quad\quad L_{2n}=\diag \left(1,-1\right)\otimes \mathbf{1}_n.
\end{align*}
\end{proposition}

\begin{proof}[{\bf Proof of Proposition~\ref{p:chpolM}}]
	The average of the product of the characteristic polynomials  over the GUE in in \eqref{FGamma} can be calculated using Grassmann integration. First we use the well-known identity
	\begin{align*}
		\intd\left(\de{\overline{\Psi}}\de{\Psi}\right)_{\mbox{ent}}
		\exp{-\langle\overline{\Psi}, M \Psi\rangle} = \det{M},
	\end{align*}
	where $M$ is $N\times N$ matrix, and $\overline{\Psi},\Psi$ are
	Grassmann variables vectors of length $N$ and
	$\left(\de{\overline{\Psi}}\de{\Psi}\right)_{\mbox{ent}}=
	\prodd\limits_{j=1}^N \de{\overline{\psi}_j}\de{\psi_j}$.
	We also write each square of determinant in the form
	\begin{align*}
		\left|\det\left(z_k-H-i\Gamma\right)\right|^2 =
		\det\left(
			\begin{array}{cc}
				z_k-H-i\Gamma & 0 \\
				0 & \overline{z}_k-H+i\Gamma
			\end{array}
		\right).
	\end{align*}
	Combining the above relations,
	\begin{align*}
		F_{\Gamma}\left(z_1,z_2,\ldots,z_n\right) =
		\\
		\MoveEqLeft[+5]
		\left\langle
			\intd
			\prodd
			\limits_{k=1}^{n}
			\left(
			\de{
			\overline{\Psi}^{\left(k\right)}}
			\de{
			\Psi^{\left(k\right)}}\right)_{\mbox{ent}}
		\exp{-\left\langle\overline{\Psi}^{\left(k\right)},
			\left(
			\begin{array}{cc}
			z_k-H-i\Gamma & 0 \\
			0 & \overline{z}_k-H+i\Gamma
			\end{array}
			\right)
			\Psi^{\left(k\right)}\right\rangle}
		\right\rangle_H.
	\end{align*}
	Now we interchange the order of integrations and perform the $\mbox{GUE}$ average first:
	\begin{align*}
		F_{\Gamma}  = &
		\left\langle
		\prodd\limits_{k=1}^n
		\exp{\left\langle\overline{\Psi}^{\left(k\right)},
			\left(
			\begin{array}{cc}
			H & 0 \\
			0 & H
			\end{array}
			\right)
			\Psi^{\left(k\right)}\right\rangle}
		\right\rangle_H
		\\
		=&
		2^{-N/2}
		\left(\frac{N}{\pi}\right)^{N^2/2}
		\intd\exp{-N\sumd\limits_{p<q}^N\left(\re h_{p,q}\right)^2
		+\left(\im h_{p,q}\right)^2 - \frac{N}{2}\sumd_{p=1}^N h_{p,p}^2}
		\\
		&\times
		\exp{\sumd\limits_{p<q}^N \re h_{p,q}		
		\sumd\limits_{k=1}^n
		\left(
		\overline{\psi}^{\left(k\right)}_{p}
		\psi^{\left(k\right)}_{q} +	
		\overline{\psi}^{\left(k\right)}_{N+p}
		\psi^{\left(k\right)}_{N+q}	
		+
		\overline{\psi}^{\left(k\right)}_{q}
		\psi^{\left(k\right)}_{p} +	
		\overline{\psi}^{\left(k\right)}_{N+q}
		\psi^{\left(k\right)}_{N+p}
		\right)}
		\\
		&\times
		\exp{i\sumd\limits_{p<q}^N \im h_{p,q}		
		\sumd\limits_{k=1}^n
		\left(
		\overline{\psi}^{\left(k\right)}_{p}
		\psi^{\left(k\right)}_{q} +	
		\overline{\psi}^{\left(k\right)}_{N+p}
		\psi^{\left(k\right)}_{N+q}	
		-
		\overline{\psi}^{\left(k\right)}_{q}
		\psi^{\left(k\right)}_{p} -	
		\overline{\psi}^{\left(k\right)}_{N+q}
		\psi^{\left(k\right)}_{N+p}
		\right)}
		\\
		&\times
		\exp{\sumd\limits_{p=1}^N h_{p,p}		
		\sumd\limits_{k=1}^n
		\left(
		\overline{\psi}^{\left(k\right)}_{p}
		\psi^{\left(k\right)}_{p} +	
		\overline{\psi}^{\left(k\right)}_{N+p}
		\psi^{\left(k\right)}_{N+q}
		\right)}\, ,
		\\
		= &
		\exp{
		\frac{1}{4N}
		\sumd\limits_{p<q}^N
		\left(
		\sumd\limits_{k=1}^n
		\left(
		\overline{\psi}^{\left(k\right)}_{p}
		\psi^{\left(k\right)}_{q} +	
		\overline{\psi}^{\left(k\right)}_{N+p}
		\psi^{\left(k\right)}_{N+q}	
		+
		\overline{\psi}^{\left(k\right)}_{q}
		\psi^{\left(k\right)}_{p} +	
		\overline{\psi}^{\left(k\right)}_{N+q}
		\psi^{\left(k\right)}_{N+p}
		\right)
		\right)^2		
		}
		\\
		&\times
		\exp{-
		\frac{1}{4N}
		\sumd\limits_{p<q}^N
		\left(
		\sumd\limits_{k=1}^n
		\left(
		\overline{\psi}^{\left(k\right)}_{p}
		\psi^{\left(k\right)}_{q} +	
		\overline{\psi}^{\left(k\right)}_{N+p}
		\psi^{\left(k\right)}_{N+q}	
		-
		\overline{\psi}^{\left(k\right)}_{q}
		\psi^{\left(k\right)}_{p} -	
		\overline{\psi}^{\left(k\right)}_{N+q}
		\psi^{\left(k\right)}_{N+p}
		\right)
		\right)^2		
		}
		\\
		&
		\times
		\exp{
		\frac{1}{2N}
		\sumd\limits_{p=1}^N
		\left(
		\sumd\limits_{k=1}^n
		\left(
		\overline{\psi}^{\left(k\right)}_{p}
		\psi^{\left(k\right)}_{p} +	
		\overline{\psi}^{\left(k\right)}_{N+p}
		\psi^{\left(k\right)}_{N+p}
		\right)
		\right)^2	
		}\\
		=&
		\exp{\frac{1}{2N}\sumd\limits_{p,q}^N
		\left(
		\sumd\limits_{k=1}^n
		\left(
		\overline{\psi}^{\left(k\right)}_{p}
		\psi^{\left(k\right)}_{q} +	
		\overline{\psi}^{\left(k\right)}_{N+p}
		\psi^{\left(k\right)}_{N+q}
		\right)
		\right)
		\left(
		\sumd\limits_{k=1}^n
		\left(		
		\overline{\psi}^{\left(k\right)}_{q}
		\psi^{\left(k\right)}_{p} +	
		\overline{\psi}^{\left(k\right)}_{N+q}
		\psi^{\left(k\right)}_{N+p}
		\right)
		\right)}.
	\end{align*}	
In the last expression one can see quartic terms in Grassmann
variables. To deal with these terms, we use the so-called Hubbard-Stratonovich transformation.
Let
	\begin{align*}
		a_{k,k'} = \sumd\limits_{j=1}^N\overline{\psi}^{\left(k\right)}_j
		\psi^{\left(k'\right)}_j, \quad
		b_{k,k'} = \sumd\limits_{j=1}^N\overline{\psi}^{\left(k\right)}_{N+j}
		\psi^{\left(k'\right)}_{N+j},\quad
		c_{k,k'} = \sumd\limits_{j=1}^N\overline{\psi}^{\left(k\right)}_j
		\psi^{\left(k'\right)}_{N+j}, \quad
		d_{k,k'} = \sumd\limits_{j=1}^N\overline{\psi}^{\left(k\right)}_{N+j}
		\psi^{\left(k'\right)}_j,
	\end{align*}
	and
	\begin{align*}
		A = \left(
			\begin{array}{c|c}
				\left\{a_{k,k'}\right\}_{k,k'=1}^n &
				\left\{c_{k,k'}\right\}_{k,k'=1}^n \\
				\hline
				\left\{d_{k,k'}\right\}_{k,k'=1}^n &
				\left\{b_{k,k'}\right\}_{k,k'=1}^n
			\end{array}
		\right).
	\end{align*}
	Then
		\begin{align*}
		\widehat{F}_{\Gamma} = \exp{-\frac{1}{2N}\Tr A^2}.
	\end{align*}
The quadratic term in the $2n\times 2n$ matrix $A$ can be linearised  at the expense of the additional integration over $2n\times 2n$ hermitian matrices $S_{2n}$ (the Hubbard-Stratonovich transformation):
	\begin{align*}
		\widehat{F}_{\Gamma} = 2^{-n} \left(\frac{N}{\pi}\right)^{2n^2}
		\int D[S_{2n}] \, \exp{-\frac{N}{2}\Tr S_{2n}^2 -i\Tr S_{2n}A}.
	\end{align*}
Now, we can integration over the Grassmann variables. We have
	\begin{align*}
		F_{\Gamma}\left(z_1,z_2,\ldots,z_n\right) = &
		\frac{1}{2^n} \left(\frac{N}{\pi}\right)^{\!\!2n^2}
		\!\!
		\int D[S_{2n}]\,
		\exp{-\frac{N}{2}\Tr S_{2n}^2} 		
		\int
		\prod
		\limits_{k=1}^{n}
		\left(
		\de{
			\overline{\Psi}^{\left(k\right)}}
		\de{
			\Psi^{\left(k\right)}}\right)_{\mbox{ent}}
	\\
	 \MoveEqLeft[7]
		\times
		\exp{-\sumd\limits_{j=1}^N
		\left(
		\sumd\limits_{k=1}^{n}z_k \overline{\psi}^{\left(k\right)}_j
		\psi^{\left(k\right)}_j+
		\sumd\limits_{k=1}^{n}\overline{z_k} \overline{\psi}^{\left(k\right)}_{N+j}
		\psi^{\left(k\right)}_{N+j}
		\right)}
	\\
	\MoveEqLeft[7]
	\times
	\exp{i\sumd\limits_{j=1}^N
		\gamma_j
		\left(
		\sumd\limits_{k=1}^{n} \overline{\psi}^{\left(k\right)}_j
		\psi^{\left(k\right)}_j-\overline{\psi}^{\left(k\right)}_{N+j}
		\psi^{\left(k\right)}_{N+j}
		\right)}
	\\
	\MoveEqLeft[7]
	\times
	\exp {-i\sumd\limits_{j=1}^N
		\left(
		\sumd\limits_{k,k'=1}^{n} s_{k',k}\overline{\psi}^{\left(k\right)}_j
		\psi^{\left(k'\right)}_j
		+s_{n+k',k}\overline{\psi}^{\left(k\right)}_j
		\psi^{\left(k'\right)}_{N+j}
		+s_{k',n+k}\overline{\psi}^{\left(k\right)}_{N+j}
		\psi^{\left(k'\right)}_j
		+s_{n+k',n+k}\overline{\psi}^{\left(k'\right)}_{N+j}
		\psi^{\left(k\right)}_{N+j}
		\right)}.
\end{align*}
By manipulating terms in the exponentials,
\begin{align*}
		F_{\Gamma}\left(z_1,z_2,\ldots,z_n\right) = &
	\frac{1}{2^n}\left(\frac{N}{\pi}\right)^{\!2n^2}
	\!\!
	\int \! D[S_{2n}] \, \exp{-\frac{N}{2}\Tr S_{2n}^2}
	\int
	\prod
	\limits_{j=1}^{N}
	\left(
	\de{\overline{\psi}_j^{\left(\cdot\right)}}
	\de{\psi_j^{\left(\cdot\right)}}\right)_{\mbox{ent}}
	\left(
	\de{\overline{\psi}_{N+j}^{\left(\cdot\right)}}
	\de{\psi_{N+j}^{\left(\cdot\right)}}\right)_{\mbox{ent}}
	\\
	&
	 \times \exp{-\left\langle
	\left(
		\begin{array}{c}
			\overline{\psi}_j^{\left(\cdot\right)} \\
			\overline{\psi}_{N+j}^{\left(\cdot\right)}
		\end{array}
	\right),
	\left[
	\left(
	\begin{array}{cc}
		Z-i\gamma_j\textbf{1}_n & 0\\
		0 & \overline{Z}-i\gamma_j\textbf{1}_n
	\end{array}
	\right)+iS_{2n}	
	\right]
	\left(
	\begin{array}{c}
		\psi_j^{\left(\cdot\right)} \\
		\psi_{N+j}^{\left(\cdot\right)}
	\end{array}
	\right)		
	\right\rangle}
	\\
	=&
	\frac{1}{2^n} \left(\frac{N}{\pi}\right)^{\!2n^2}
	\!\!
	\int D [S_{2n}]\,  
	\exp{-\frac{N}{2}\Tr S_{2n}^2}
	\prodd\limits_{j=1}^N\det\left(Z_{2n}+iS-i\gamma_jL_{2n}\right).
	\end{align*}	
	Now, recalling that $\gamma_j=0$ for $j=M+1, \ldots, N$
	we obtain the statement of the Proposition.	
\end{proof}

\section[\appendixname~\thesection]{Various asymptotic regimes for Laguerre polynomials}
Asymptotic behaviour of the Laguerre polynomials $L_{N-k}^{\left(\alpha\right)}\left(-NY^2\right)$ in the limit when $N\to\infty$ and $k$ and $\alpha$ are fixed depends on the scale of the variable $Y>0$ compared to $N$. For our investigation we need two scales: (i) $YN=y>0$ is fixed and (ii) $YN\gg 1$. In both cases the desired approximations can be obtained from the integral representation \eqref{Lagint} which we rewrite as
%
\begin{align}\label{Lagasy1}
		L_{N-k}^{(\alpha)}\left(-NY^2\right) =
		\frac{2 N^{N-k+1}}{\left(N-k\right)!}\frac{e^{-NY^2}}{Y^{\alpha}}
	\int \limits_{0}^{\infty} \tau^{-2k+\alpha+1}e^{-N(\tau^2-2\ln{\tau})} I_{\alpha}(2\tau Y\!N)\, d\tau, \quad \quad Y>0.
	\end{align}
We start with simpler case of $YN=y>0$ being fixed in the limit $N\to\infty$. In this case significant contributions to the integral in \eqref{Lagasy1} are coming from a neighbourhood of the point $\tau=1$ which is the point of minimum the function $\tau^2-2\ln{\tau}$ inside the interval of integration. Straightforward evaluation of the integral by the Laplace method together with the Stirling approximation $(N-k)!\sim \sqrt{2\pi} e^{-N}N^{N-k+1/2}$ yields that
\begin{align}\label{Lagasy2}
 L_{N-k}^{(\alpha)}\left(-\frac{y^2}{N}\right)\sim \frac{N^{\alpha}}{y^{\alpha}} I_{\alpha}(2y)\, , \quad\quad  \quad y>0.
\end{align}
In the other regime of interest for us, $YN\gg 1$, one can use the following asymptotic expansion for the modified Bessel function $I_{\alpha}(z)$ (see e.g.
formula 5.11.10 in \cite{Lebedevbook}):
\begin{align}
I_{\alpha}(z)=\frac{e^z}{\sqrt{2\pi z}}\sum_{p=0}^n\frac{(-1)^p}{(2z)^p}A^{\left(\alpha\right)}_p+O\left(|z|^{-n-1}\right), \quad A^{\left(\alpha\right)}_p=\frac{\Gamma(\alpha+p+1/2)}{\Gamma(\alpha-p+1/2)}.
\end{align}
It reduces the asymptotic analysis of $L_{N-k}^{(\alpha)}\left(-N Y^2 \right)$ to analysis of the following expression:
\begin{align}\label{Lagasy3}
 \frac{e^{-NY^2}}{2Y^{\alpha}\sqrt{\pi YN}} \int \limits_{0}^{\infty}\tau^{-2k+\alpha+1/2}e^{-N \mathcal{L} (\tau)}\sum_{p}\frac{(-1)^p}{(4YN)^p}A^{\left(\alpha\right)}_p\,  d \tau, \quad \mathcal{L}(\tau)=\tau^2-2\ln{\tau}-2\tau Y.
\end{align}
In this case significant contributions to the integral in \eqref{Lagasy3} are coming from a neighbourhood of
the point $\tau=\tau_*(Y)$ which is the point of minimum the function $\mathcal{L}(\tau)$ inside the interval of integration.
\begin{align}
 \tau_*(Y)=\frac{Y+\sqrt{Y^2+4}}{2}=\frac{1}{r_*(Y)}, \quad r_*(Y)=\frac{\sqrt{Y^2+4}-Y}{2}>0\, ,
\end{align}
Using the relations $\tau_*(Y)=r_*(Y)+Y$ and  $1+r_*(Y)^2=r_*(Y) \left(Y^2+4\right)^{1/2}$ we find that
\begin{align}
\mathcal{L}(\tau_*(Y))=1+2\ln{r_*(Y)}-Y(r_*(Y)+Y), \quad 
\mathcal{L}^{\prime\prime} (\tau_*) =
2r_*(Y)\left(Y^2+4\right)^{1/2}\, .
\end{align}
Expanding the integrand in the standard way around $\tau=\tau_*(Y)$ and collecting the leading and subleading order terms we get asymptotic expressions for Laguerre polynomials with the precision sufficient for our purposes:
\begin{align}\label{AsyLagg}
		L_{N-k}^{(\alpha)}(-NY^2) \!=\!
\begin{cases}
\displaystyle{
			\frac{e^{NYr_*(Y)}}{\sqrt{2\pi N}}\frac{ r_*(Y)^{-2(N-k)-\alpha-1}}		{Y^{\alpha+\frac{1}{2}}\left(Y^2+4\right)^{1/4}}\left[1\!+\!O\left(\!\frac{1}{N}\!\right)\!\right],
			} &  Y=O(1), \\[4ex]
\displaystyle{
		\frac{e^{NYr_*(Y)}}{\sqrt{2\pi N}}\frac{ r_*(Y)^{-2(N-k)-\alpha-1}}		{Y^{\alpha+\frac{1}{2}}\left(Y^2+4\right)^{1/4}}
		\left[1\!-\!\frac{(4\alpha^2-1)r_*(Y)}{16YN}\!+\!O\!\left(\!\frac{1}{N}\!\right)\!\right],\!\!
		}& Y\ll 1 \ll NY.
\end{cases}
	\end{align}


\end{document}